\documentclass[preprint,preprintnumbers,prl,amsmath,amssymb,superscriptaddress,floatfix]{revtex4}
\usepackage{graphicx}% Include figure files
\usepackage{bm}% bold math
\usepackage{amssymb}
\usepackage{units}
\usepackage{microtype}
\usepackage{ccaption}
\usepackage{ragged2e}
  \captionstyle{\justifying}
\begin{document}

\title{Quantifying octahedral rotations in strained perovskite oxide films}

\author{S.\ J.\ May}
\email{smay@drexel.edu}
\affiliation{
Materials Science Division, Argonne National Laboratory, Argonne, IL 60439, USA\\}
\affiliation{
Department of Materials Science and Engineering, Drexel University, Philadelphia, PA 19104, USA\\}
\author{J.-W.\ Kim}
\affiliation{
Advanced Photon Source, Argonne National Laboratory, Argonne, IL 60439, USA\\}
\author{J.\ M.\ Rondinelli}
\affiliation{
Materials Department, University of California, Santa Barbara, CA 93106, USA\\}
\author{E.\ Karapetrova}
\affiliation{
Advanced Photon Source, Argonne National Laboratory, Argonne, IL 60439, USA\\}
\author{N.\ A.\ Spaldin}
\affiliation{
Materials Department, University of California, Santa Barbara, CA 93106, USA\\}
\author{A.\ Bhattacharya}
\affiliation{
Materials Science Division, Argonne National Laboratory, Argonne, IL 60439, USA\\}
\affiliation{
Center for Nanoscale Materials, Argonne National Laboratory, Argonne, IL 60439, USA\\}
\author{P.\ J.\ Ryan}
\affiliation{
Advanced Photon Source, Argonne National Laboratory, Argonne, IL 60439, USA\\}

\date{\today}

%\maketitle

\begin{abstract}
Epitaxial strain is a proven route to enhancing the properties of complex oxides, however, the details of how the atomic structure accommodates strain are poorly understood due to the difficulty of measuring the oxygen positions in thin films.  We present a general methodology for determining the atomic structure of strained oxide films via x-ray diffraction, which we demonstrate using LaNiO$_3$ films.  The oxygen octahedral rotations and distortions have been quantified by comparing the intensities of half-order Bragg peaks, arising from the two unit cell periodicity of the octahedral rotations, with the calculated structure factor.  Combining \textit{ab initio} density functional calculations with these experimental results, we determine systematically how strain modifies the atomic structure of this functional oxide.
\end{abstract}

\maketitle

\begin{figure}
\includegraphics[width=3.6 in]{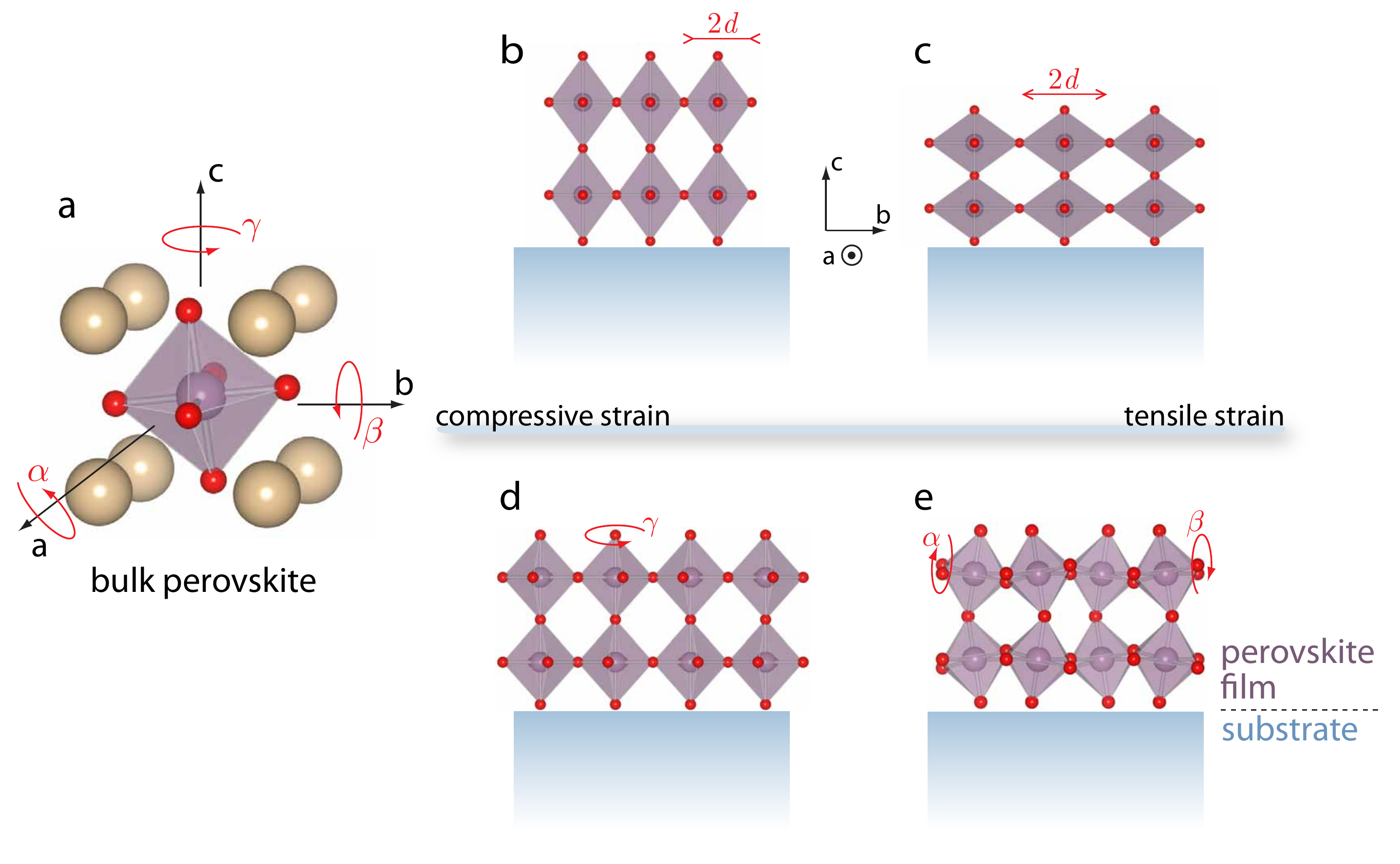}
\caption{The available structural distortions afforded in perovskite oxides through heteroepitaxial thin film growth.  a, Rotations of the oxygen octahedra are pervasive in bulk perovskites and commonly decomposed into components $\alpha, \beta$ and $\gamma$  about each of the Cartesian axes.  In coherently strained perovskite films, the $B$O$_6$ octahedra can distort through a contraction (\textbf{b}) or elongation (\textbf{c}) of the in-plane $B$-O bond lengths ($d$) under compressive or tensile strain, respectively.  Simultaneously, the octahedra can accommodate the substrate-induced in-plane lattice parameters by rotating perpendicular to the substrate (\textbf{d}), and/or rotating along the axes parallel to the substrate plane (\textbf{e}).}
\label{fig:tilts}
\end{figure}

Strain engineering is an appealing route to controlling electronic \cite{Ward09}, ferroic \cite{Tokura00,Schlom07,Ramesh07}, and structural properties in complex oxide heterostructures \cite{He04,Ramesh09}.  The sensitivity to strain arises from the strong electron-, spin-, and orbital-lattice coupling inherent to transition metal oxides\cite{Torrance92,Chmaissem01,Mizokawa99}.  In $AB$O$_3$ perovskites (Fig.\ \ref{fig:tilts}a), the flexible corner-shared $B$O$_6$ octahedral network allows epitaxial strain to be accommodated by either bond-length distortions (Fig.\ \ref{fig:tilts}b,c), and/or by changes in the magnitude and phase of the octahedral rotations (Fig.\ \ref{fig:tilts}d,e).  Previous work has demonstrated how strain-based manipulation of bond-length distortions can be used to dramatically enhance ferroelectricity \cite{Choi04,Haeni04}.  A less explored, yet perhaps equally productive \cite{Zayak06,Hatt08} strategy is the control of {\it bond angles} to manipulate magnetism, metal-insulator transitions, and superconductivity through tuning of superexchange pathways and electronic bandwidths.  However, the current understanding of strain--bond angle relationships is limited, in part due to the experimental challenge of measuring oxygen positions in thin films \cite{Xie08,Woodward05,Jia09}.  This limited understanding of how strain alters both bond lengths and angles is a clear obstacle to engineering tailored functionalities into oxide films and heterostructures.

In perovskite oxides, octahedral rotations offset the oxygen atoms from the face-centered positions effectively doubling the pseudocubic unit cell, producing a distinctive set of half-order Bragg peaks depending on the tilt pattern \cite{Glazer75}.  The octahedra can rotate about the pseudocubic [1 0 0] ($a$), [0 1 0] ($b$), and/or [0 0 1] ($c$) directions \cite{Glazer72,Woodward97a,Woodward97b}; these rotation angles are referred to as $\alpha$, $\beta$, and $\gamma$, respectively, and are depicted in Fig.~\ref{fig:tilts}a.  To describe the phase of the octahedral rotations along each axis, we employ the Glazer notation, in which a superscript is appended to each axis to indicate whether neighboring octahedra rotate in-phase ($^+$), out-of-phase ($^-$) or if rotations are absent ($^0$) \cite{Glazer72}.  In the absence of cation-site displacements, the allowed and forbidden half-order diffraction peaks are a direct fingerprint of the octahedral rotations in perovskite oxides.

While half-order peaks have been employed previously to identify crystallographic space groups of perovskite films \cite{He05}, in this work we demonstrate how these Bragg peaks can be used to \textit{quantify the octahedral rotations and bond angles} of strained perovskite oxide films.  The oxygen positions, and thus the bond angles and lengths, are obtained by comparing the intensities of a systematic series of half-order Bragg peaks with the calculated structure factor of the oxygen octahedra.  Combining the diffraction results with density functional calculations, we examine the evolution of the rotation angles with strain for a model functional oxide.  The relative simplicity of the experimental approach, requiring only high flux x-ray diffraction and the calculation of structure factors, makes this a promising strategy for determining the full atomic structure of a broad range of strained perovskite films.

We chose to study LaNiO$_3$ (LNO) as an example because it exhibits octahedral rotations along each Cartesian axis.  Bulk LNO has an $a^-a^-a^-$ rotation pattern ($\alpha = \beta = \gamma = 5.2^\circ$) with a $R\bar{3}c$ symmetry and a pseudocubic lattice parameter of 3.838 \AA~ \cite{Garcia92}.  This results in uniform Ni-O-Ni bond angles of 165.2$^\circ$ and Ni-O bond lengths of 1.935 \AA.  The isotropic nature of bulk LNO is ideally suited to exploring how strain alters bond lengths and angles in an anisotropic manner.  Additionally, LNO has garnered interest due to its novel properties such as electric-field control of conductivity \cite{Scherwitzl09}, correlated Fermi liquid behavior \cite{Eguchi09}, and recent predictions of superconductivity \cite{Hansmann09}.

Our epitaxial LNO films were deposited on SrTiO$_3$ ($a$ = 3.905 \AA) and LaAlO$_3$ ($a$ = 3.795 \AA) substrates by molecular beam epitaxy \cite{May09} in order to probe how both tensile (1.7 $\%$ on STO) and compressive (-1.1\% on LAO) strain are accommodated by changes in the bond angles and lengths in the perovskite structure. The films grown on SrTiO$_3$ (STO) and LaAlO$_3$ (LAO) are 173 \AA~and 95 \AA~thick, respectively, and coherently strained with $c$-axis parameters of 3.807 and 3.895~{\AA} as confirmed through reciprocal space maps and (00$L$) scans.  All x-ray diffraction measurements were performed at room temperature.

In addition to exploring how strain modifies the octahedral distortions, growing on the two substrates, STO and LAO, enables us to investigate how the substrates' structural properties interact with the octahedral rotations of the LNO films through heteroepitaxy.  For example, STO is cubic at room temperature and lacks octahedral rotations, whereas LAO is rhombohedral with the $a^-a^-a^-$ rotation pattern.  Thus, by measuring the structural distortions in the LNO/STO films, we can determine whether octahedral rotations are suppressed when growing on a cubic substrate.   Similar analysis of the distortions in the LNO films on LAO enables us to determine if the rotational pattern present in the substrate couples to that found in the film.

\begin{figure}
\includegraphics[width=3.6 in]{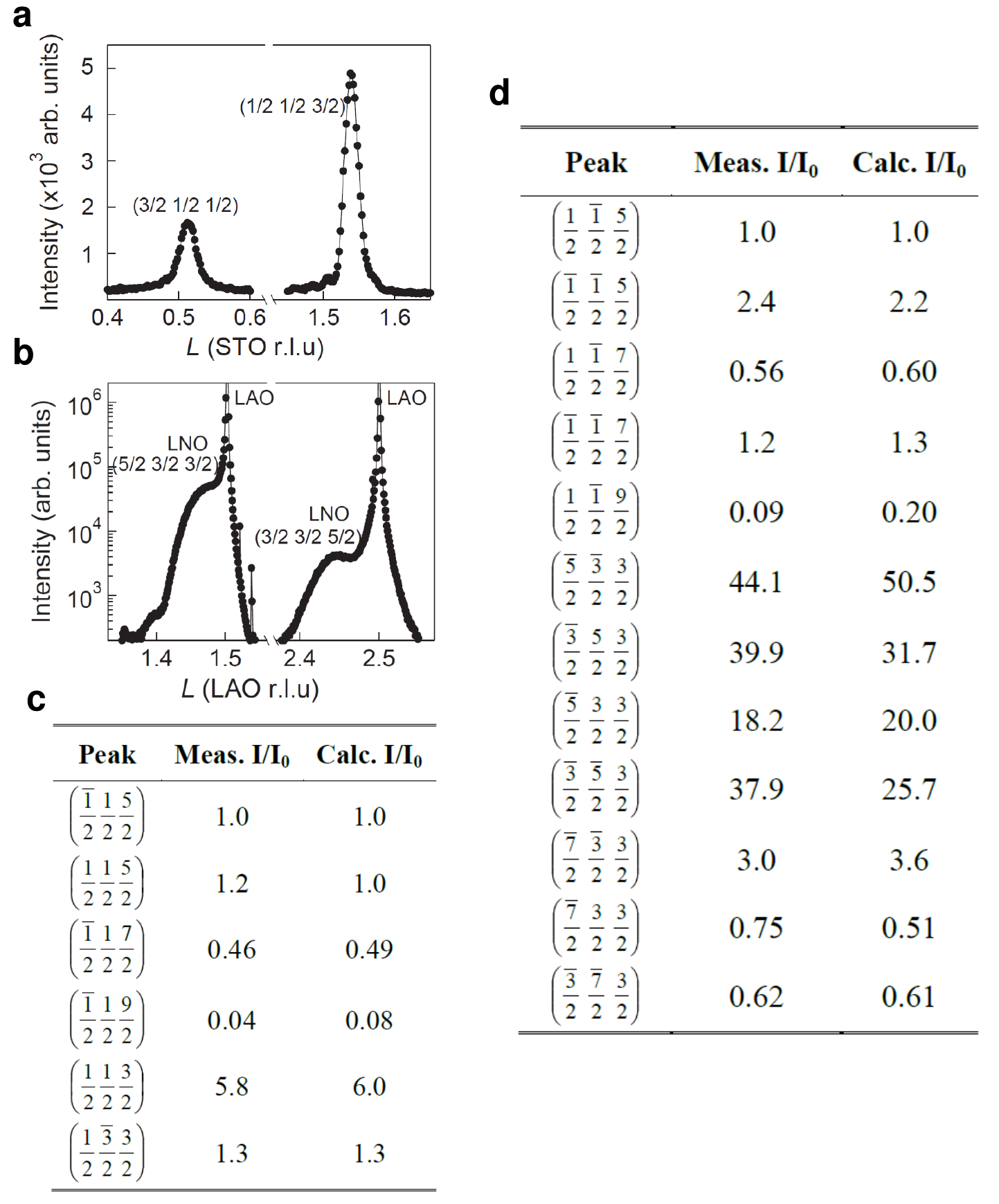}
\caption{Half-order peaks of LNO films grown on STO and LAO.  \textbf{a,} In LNO on STO, the $h=k$ diffraction peaks are more intense than those with $h\neq k$, as $\alpha > \gamma$.  \textbf{b,} The opposite intensity pattern is observed for the LNO on LAO, as $\alpha < \gamma$.  Intense half-order peaks from octahedral rotations within the LAO substrate are also present; this intensity from the substrate is less than 10$\%$ of the film peak intensity.  \textbf{c,} Comparison of the measured and calculated peak intensities for LNO on STO for $\alpha = \beta = 7.1^{\circ}$ and $\gamma=0.3^{\circ}$.  \textbf{d,} Comparison of the peak intensities for LNO on LAO for $\alpha = \beta = 1.7^{\circ}$ and $\gamma=7.9^{\circ}$, with a domain ratio of 1:2.15:0:0.}
\label{fig:xrd}
\end{figure}

The rotation pattern of the LNO films is identified by observing the presence and absence of specific half-order Bragg peaks.  We find half-order peaks in the films grown on both substrates (Fig.\ \ref{fig:xrd}a,b), which confirms that heteroepitaxial growth on a cubic substrate does not suppress octahedral rotations by forcing atomic registry between the film and the substrate; the epitaxial relationship is not simply ``cube-on-cube.''  Both films exhibit half-order peaks when the following conditions are met: 1) $h$, $k$, and $l$ are equal to $n$/2, where $n$ is an odd integer and 2) $h=k\neq l$, $k=l\neq h$, and $h=l\neq k$.  Peaks are not observed where $h=k=l$, such as ($\nicefrac{1}{2}$ $\nicefrac{1}{2}$ $\nicefrac{1}{2}$), or where one of the indices is equal to an integer, such as ($\nicefrac{1}{2}$ 1 $\nicefrac{3}{2}$).  This reciprocal lattice indicates the presence of an $a^-a^-c^-$ tilt pattern within both films \cite{Glazer75}, with $\alpha$ = $\beta$ due to the biaxial strain state of the films ($a$ = $b$).

The intensities of the half-order peaks can be compared to calculated peak intensities in order to quantify the octahedral rotation angles.  The calculated peak intensities are determined from Eq. (\ref{eq:one}):
\begin{equation}
  I=I_0 \frac{1}{\sin(\eta)} \frac{1}{\sin(2\Theta)} (\displaystyle\sum_{j=1}^{4}D_j|F_{hkl}|^2),
  \label{eq:one}
\end{equation}
where $I_0$ is the incident photon flux, 1/${\sin(\eta)}$ is a correction for the beam footprint, 1/${\sin(2\Theta)}$ is the Lorentz polarization correction, $F_{hkl}$ is the structure factor for the oxygen atoms, and $D_j$ is the relative volume fraction of structural domains discussed below.  The incident photon angle is $\eta$, while the scattering angle is $\Theta$.  The structure factor for each Bragg peak is obtained from,
\begin{equation}
  F_{hkl}= f_{O^{2-}} \displaystyle\sum_{n=1}^{24} \exp [2\pi i(hu_n + kv_n + lw_n)],
  \label{eq:two}
\end{equation}
where the position of the $n^{th}$ oxygen atom within the unit cell is given by ($u$, $v$, $w$), and $f_{O^{2-}}$ is the O$^{2-}$ form factor  \cite{Hovestreydt}.  The La and Ni atoms are assumed to reside on the ideal corner and body-centered positions, respectively, and thus we do not consider a rhombohedral distortion of the La positions.  Given this assumption, only the oxygen atoms contribute to the half-order peak intensities.  We account for the octahedral rotations by doubling the unit cell in each direction, giving rise to a non-primitive unit cell consisting of 24 oxygen positions (8 octahedra).  The oxygen positions are displaced from the ideal face-centered positions by the appropriate rotation angle.  To calculate the intensity of a given peak, the relative volume of all possible structural domains ($D_j$) is needed.  When calculating the structure factor, these domains determine whether the octahedron nearest to the chosen origin rotates clockwise or counterclockwise about each axis.  Within this tilt pattern there are four domains that give rise to different structure factors: one where $\alpha$, $\beta$, and $\gamma$ are all positive, and three where two of the angles are positive and the third is negative.  The domain volume fractions are obtained from a set of symmetrically equivalent half-order peaks with fixed $l$, for instance ($\pm\nicefrac{1}{2}$ $\pm\nicefrac{1}{2}$ $\nicefrac{3}{2}$), the relative intensities of which depend directly on the domain populations.  We employ standard non-linear regression analysis to determine the optimal values of $\alpha$ and $\gamma$ and their respective errors.

The films grown on cubic STO and rhombohedral LAO substrates exhibit different domain populations.  In the LNO/STO film, we find an equal volume fraction of the four possible domains, indicating that the four domains are energetically equivalent when LNO is grown on a cubic substrate.  In contrast the LNO/LAO film shows unequal domain populations between only two domains with an approximate 1:2 ratio.  To explain this difference, we note that our LAO substrates contain extensive twinning, which we observed as two distinct peaks in $\omega$-scans rocked about a (0 0 $l$) peak.  For our LAO substrate the ratio of the two peaks is also 1:2 suggesting that the octahedral rotations and domain pattern of the substrate are transferred to the film. Since the rhombohedral distortion determines the trigonal axis about which the octahedra rotate in bulk LAO, this result demonstrates that substrates may be used to texture octahedral rotations by choosing appropriate substrate orientations. Heteroepitaxy therefore affords an additional mechanism by which epitaxial thin film growth can be used to engineer the physical properties of complex oxides.  Additionally, imprinting of rotational behavior may play an important role in the interfacial properties of oxides.

\begin{figure}
\includegraphics[width=3.6 in]{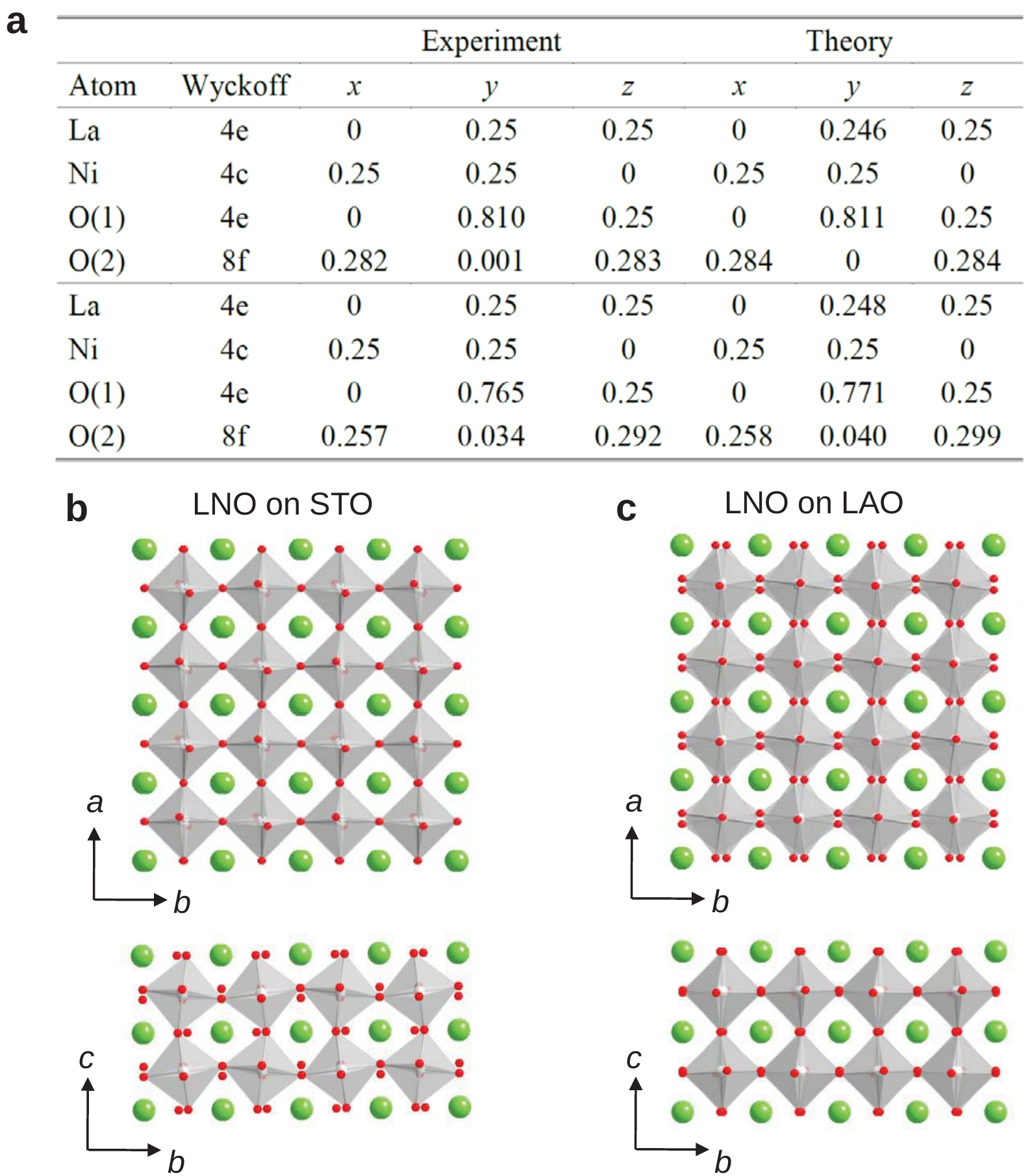}
\caption{Atomic structure of LNO grown on STO and LAO as determined by x-ray diffraction (Experiment) and density functional theory (Theory).  In \textbf{a}), the top data set corresponds to LNO/STO, while the bottom data set corresponds to LNO/LAO.  The space group is $C 2/c$ (\#15), with $a$ = 9.406 \AA, $b=c$ = 5.523 \AA, and a monoclinic angle of 125.95$^\circ$ for LNO/STO and $a$ = 9.460 \AA, $b=c$ = 5.367 \AA, and a monoclinic angle of 124.57$^\circ$ for LNO/LAO.  The relation of the monoclinic unit cell ($a$,$b$,$c$) to the ideal perovskite ($a_p$,$b_p$,$c_p$) is $a_p$ = $c$/$\sqrt{2}$, $b_p$ = $b$/$\sqrt{2}$, and $c_p =\sqrt{6}a$.  \textbf{b,} Tensile strain from the STO substrate reduces the $\gamma$ rotation, while increasing the $\alpha$ and $\beta$ rotations.  \textbf{c,} Compressive strain induced from the LAO substrate has the opposite effect: increasing the $\gamma$ rotation, while decreasing the $\alpha$ and $\beta$ rotations.}
\label{fig:table}
\end{figure}

For LNO on STO, we obtain $\alpha = \beta = 7.1 \pm 0.2^{\circ}$ and $\gamma=0.3 \pm 0.7^{\circ}$.  The measured and calculated intensities are given in Fig.~\ref{fig:xrd}c.  For LNO on LAO, we obtain $\alpha = \beta = 1.7 \pm 0.2^{\circ}$ and $\gamma=7.9 \pm 0.9^{\circ}$; the peak intensities are provided in Fig.~\ref{fig:xrd}d.  Knowing the tilt angles, and thus the oxygen positions, the atomic structures are obtained (Fig.~\ref{fig:table}).

As anticipated, the epitaxial strain strongly modifies the in-plane (equatorial) Ni-O bond lengths ($d_{Ni-O}$) due to coherency across the heterointerface.  The in-plane $d_{Ni-O}$ decreases systematically from LNO/STO (1.968 $\pm$ 0.002 \AA) to bulk LNO (1.935 \AA) to LNO/LAO (1.916 $\pm$ 0.005 \AA) as the $a$- and $b$-axis parameters of the substrate are reduced. The measured out-of-plane (apical) $d_{Ni-O}$, however, are less sensitive to the strain; we obtain 1.933 $\pm$ 0.002 \AA~ and 1.950 $\pm$ 0.002 \AA~ for LNO/STO and LNO/LAO, respectively.  (Supplementary Table S1 provides a full summary of our experimentally determined bond lengths and angles.) Interestingly we find that bi-axial strain only weakly modifies the in-plane Ni-O-Ni bond angle ($\theta$) of $165.8 \pm 0.5^{\circ}$ and $164.0 \pm 2.0^{\circ}$ for LNO/STO and LNO/LAO, respectively.  In contrast, along the out-of-plane direction $\theta$ is highly sensitive to the substrate-induced strain; this bond angle decreases from $175.2 \pm 0.6^{\circ}$ in LNO/LAO to $159.9 \pm 0.6^{\circ}$ in LNO/STO.  We note that both the $\alpha$ (or $\beta$) and $\gamma$ angles contribute to the in-plane bond angles and since we find that empirically the average of these angles is robust to strain, the insensitivity of the equatorial bond angle is not surprising, albeit counterintuitive.

\begin{figure}
\includegraphics[width=2.5in]{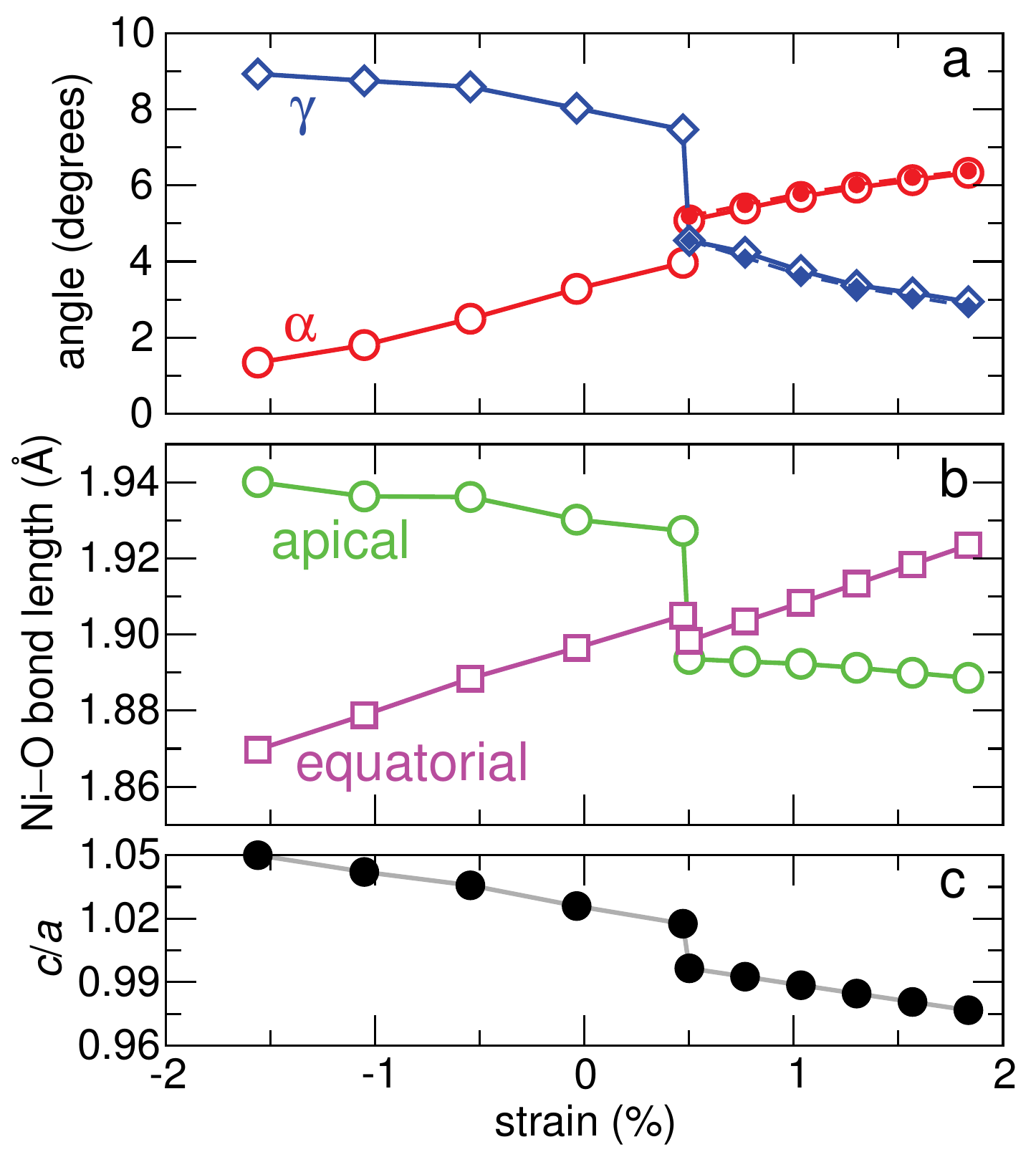}
\caption{The DFT calculated rotation angles $\alpha$ and
$\gamma$ (\textbf{a}), Ni--O bond lengths (\textbf{b}) and the $c/a$ axial ratio (\textbf{c}) for the
monoclinic nickelate structure as a function of strain (open symbols).
A sharp first-order transition induced by heteroepitaxy occurs at
approximately $c/a=1$.
The solid symbols (broken lines) in (\textbf{a}) indicate the bond angles
associated with a lower symmetry structure described in the text.}
\label{fig:dft}
\end{figure}

To investigate the origin of these effects and to determine the
change in octahedral rotations for intermediate strain values,
we perform density functional theory (DFT) calculations within the
local-spin density approximation (LSDA) plus Hubbard $U$ method
($U_{\rm eff}=U-J = 3$~eV) as implemented in the Vienna
{\it ab initio} Simulation Package ({\sc vasp}) \cite{Anisimov97,Dudarev98,Kresse96,Kresse99}.
We first  optimize the internal degrees
of freedom for bulk rhombohedral LNO (10-atom unit cell) at the
experimental lattice parameters to confirm that we accurately describe the
electronic and atomic structure.
Consistent with the experimental data, our optimized structure is indeed
metallic and consists of equal and alternating phase rotations
of the NiO$_6$ octahedra about each Cartesian axis with
$\alpha=\beta=\gamma = 5.76^\circ$, in good agreement with
the 1.5~K structural measurement of 5.37$^\circ$ \cite{Garcia92}.
We next carry out structural optimizations for the two homoepitaxially
strained films using our experimentally determined lattice parameters.
The substrates are not explicitly simulated in our calculations, but
the symmetry reduction imposed by enforcing equal in-plane lattice
parameters on the bulk  $R\bar{3}c$ structure is included
by optimizing the internal degrees of freedom within a monoclinic
(20-atom) unit cell (space group  $C2/c$).
The excellent consistency between our experimental and theoretical \
structures is  apparent in the structural data obtained for
the nickelate films grown on STO and LAO as shown in
Fig.~\ref{fig:table}a.
To quantify the comparison between the two methods, we
use a symmetry adapted mode analysis
\cite{Stokes/Hatch:1988,Orobengoa/Perez-Mato_et_al:2009} and take scalar
products between the DFT and experimentally determined vectors $\xi$
that contain the full irreducible representations of the atomic displacement
patterns  of the strained nickelate films about the bulk
rhombohedral structure.
We find that $\xi(\textrm{\sc dft})\cdot\xi(\textrm{\sc exp})\sim0.99$
for LNO on both LAO and STO which indicates excellent
accuracy between the two structure determination methods.
We note that the DFT ground state indicates an additional energy
lowering distortion through small antiparallel displacements of
$\sim 0.01$~{\AA} for the La atoms along the [110] directions.
Such displacements would produce weak but measurable
($\nicefrac{1}{2}$ $\nicefrac{1}{2}$ $\nicefrac{1}{2}$) and
($\nicefrac{3}{2}$ $\nicefrac{3}{2}$ $\nicefrac{3}{2}$) peaks, which however
were not observed in the diffraction measurements, and likely persists in
the DFT calculations due to the common exchange-correlation
functional underestimation of the equilibrium volume.
We next extend our analysis to intermediate strain values to investigate
the transition region between the experimentally obtained strain values.
In this section, we carry out structural relaxations about the theoretical LSDA$+U$ equilibrium volume rather than at the experimental lattice parameters reported previously.
We explore bi-axial strain states by fixing the in-plane lattice constants at each strain value, and performing full structural optimizations of the internal atomic coordinates and $c$-axis parameter about the LSDA$+U$ reference structure.
In Fig. \ref{fig:dft}a we show that over the strain range investigated the
NiO$_6$ octahedra always rotate out-of-phase about each axis with the
$a^-a^-c^-$ rotation pattern with only changes in their magnitude about each axis.
In agreement with the experimental data, we find that compressive (tensile)
strain predominately induces rotations of the NiO$_6$ octahedra about the
$c$-axis (in-plane axes).
We attribute this effect to the epitaxial constraint that forces
coherency between the film and the
substrate, which contracts (elongates) the equatorial bond lengths and
elongates (contracts) the apical bonds under compressive (tensile)
strain (Fig.\ \ref{fig:dft}b).
The $a^-a^-c^-$ octahedral rotation pattern is able to accommodate the
bond length distortions by increasing the NiO$_6$ rotation about the
$c$-axis at the expense of the in-plane rotations.
For the tensile case the amount of in-plane rotation decreases accordingly
to sustain the bond length elongation in-plane.
We find that there is an additional structural phase that is
approximately 2~meV lower in energy for small tensile strains and
shows an intriguing small charge disproportionation (CDP).
The deviations in rotation angles between the two phases shown
in Fig. \ref{fig:dft}a are negligible, and since our room temperature
experiments are likely unable to access this phase, we do not discuss it
further.

The transition region which captures the crossover from mainly rotations
about the $c$-axis to the in-plane axes occurs near 0.50\% strain or
when the axial ratio  $c/a$ approaches unity (Fig.\ \ref{fig:dft}c).
At this strain value, standard crystallographic distortion parameters
that measure deviations in ideal six-fold coordination through the
apical and equatorial bond lengths and the Ni--O--Ni bonds
all show a sharp discontinuity and indicate almost ideal octahedra
 \cite{Robinson71}.
In the absence of the small CDP phase, we identify this abrupt, first-order phase transition to be {\it isosymmetric} \cite{Christy:95}
since the atomic structure remains monoclinic---the full
symmetry of the $C2/c$ space group is retained and the same
Wyckoff positions are occupied---across the transition.
The origin for first-order phase transition about $c/a=1$ can be
understood on geometric grounds (Supplementary Fig.\ S5): when the ideal NiO$_6$ octahedra
is recovered at 0.5\% strain, the bi-axial lattice distortion imposed on
the film by the substrate leads to a tiling of corner-connected octahedra
that is incompatible with size of the unit cell.
Recently a strain-induced isosymmetric phase transition, with the
axial ratio being the order parameter has also been
reported in insulating thin films of the rhombohedral
antiferromagnetic and ferroelectric BiFeO$_3$ \cite{Ramesh09}.
Since LNO and BiFeO$_3$ have distinctly different physical
properties (LNO is metallic and non-polar), we suggest this intriguing
strain-induced isosymmetric phase transition may be a universal
feature of epitaxially strained rhombohedral oxide films with the
$a^-a^-c^-$ rotation pattern.
In principle such strain-induced changes to the rotation patterns
can be used to tailor the bandwidth anisotropy, and thus macroscopic
properties such as conductivity and magnetism in the thin film.
In the case of the large bandwidth compound LNO, however, our
DFT calculations (Supplementary Fig.\ S6) and experiments indicate that the strain induced by
the STO and LAO substrates is insufficient to produce
dramatic changes in the electronic structure.

In summary, we have demonstrated a general strategy for quantifying oxygen octahedral rotations in strained complex oxide films.  We have determined the atomic structure of strained LaNiO$_3$, in which the NiO$_6$ octahedra are distorted and rotated to accommodate strain.  It is anticipated that the future application of this method will enhance fundamental studies of strain accommodation and relaxation in perovskite oxides, allow for more direct comparisons between bulk and thin film materials, and help elucidate the physical mechanisms of novel properties at oxide interfaces.

\section*{Methods}

X-ray diffraction measurements were performed at the Advanced Photon Source on Sectors 33-BM and 6-ID using four-circle and six-circle Huber diffractometers.  When analyzing the peak intensities, we choose to use the peak height as a measure of intensity.  Scans along $h$ and $l$ confirm that the peak widths are independent of their location on the reciprocal lattice.  The measured intensity is determined by normalizing for the incident intensity and subtracting background intensity.  The calculated intensity is determined using Eqns. (1) and (2).  The corrections for the beam footprint, Lorentz polarization factor, and the $q$ dependence of the O$^{2-}$ form factor are taken into account in the calculated intensity.  The adjustable parameters are the rotation angles, $\alpha$ and $\gamma$.  In the case of LNO/LAO, we did not attempt to remove any diffuse background from the LAO substrate, which exhibits strong half-order peaks.  We estimate the magnitude of this substrate-induced background to be less than 10$\%$ of the intensity of the film peak.  This value is less than the average agreement between the measured and calculated peak intensities.  As such, we do not believe this introduces a significant error into our determination of $\alpha$ and $\gamma$.

We carry out first-principles density functional theory (DFT) calculations within the local-spin density approximation (LSDA) plus Hubbard $U$ method \cite{Anisimov97} as implemented in the Vienna \textit{ab initio} Simulation Package (VASP) \cite{Kresse96,Kresse99}.  We follow the Dudarev approach \cite{Dudarev98} and include an effective Hubbard term $U_{eff}  = U - J$ of 3 eV to accurately treat the correlated Ni 3$d$ orbitals.  The core and valence electrons are treated with the projector augmented wave (PAW) method \cite{Bleochl94} with the following valence electron configurations: $5p^65d^16s^2$ (La), $3p^63d^94s^1$ (Ni), and $2s^22p^4$ (O).  The Brillouin zone integrations are performed with a Gaussian smearing of 0.05 eV over a 5x5x5 Monkhorst-Pack $k$-point mesh \cite{Monkhorst76} centered at $\Gamma$, and a 450 eV plane-wave cutoff.  For structural relaxations we also used a Gaussian broadening technique of 0.05 eV and relaxed the ions until the Hellmann-Feynman forces are less than 1 meV \AA$^{-1}$.  In all calculations ferromagnetic spin order is enforced, and for our choice of $U_{eff}$, we accurately reproduce the enhanced paramagnetic spin moment \cite{Sreedhar92}.

\section*{Acknowledgements}

Work supported by UChicago Argonne, LLC, Operator of Argonne National Laboratory (ÒArgonneÓ). Work at Argonne and use the Advanced Photon Source at Argonne was supported by the U. S. Department of Energy, Office of Science, Office of Basic Energy Sciences, under Contract No. DE-AC02-06CH11357.  We acknowledge use of the oxide-MBE at the Center for Nanoscale Materials.  We acknowledge support from NDSEG (JMR) the NSF under grant no.\ DMR 0940420 (NAS) and discussions with the Army Research Office MURI program (award  no.\ W911-NF-09-1-0398). Portions of this work made use of the SGI Altix {\sc cobalt} system at the National Center for Supercomputing Applications under grant no.\ TG-DMR-050002S, and the CNSI Computer Facilities at UC Santa Barbara under NSF award no.\ CHE-0321368. Correspondence regarding the diffraction measurements should be made to P.R. (pryan@aps.anl.gov).  Correspondence regarding the DFT calculations should be made to J.M.R. (rondo@mrl.ucsb.edu).

%\bibliography{tilts,theory}

\begin{thebibliography}{38}
\expandafter\ifx\csname natexlab\endcsname\relax\def\natexlab#1{#1}\fi
\expandafter\ifx\csname bibnamefont\endcsname\relax
  \def\bibnamefont#1{#1}\fi
\expandafter\ifx\csname bibfnamefont\endcsname\relax
  \def\bibfnamefont#1{#1}\fi
\expandafter\ifx\csname citenamefont\endcsname\relax
  \def\citenamefont#1{#1}\fi
\expandafter\ifx\csname url\endcsname\relax
  \def\url#1{\texttt{#1}}\fi
\expandafter\ifx\csname urlprefix\endcsname\relax\def\urlprefix{URL }\fi
\providecommand{\bibinfo}[2]{#2}
\providecommand{\eprint}[2][]{\url{#2}}

\bibitem[{\citenamefont{Ward et~al.}(2009)\citenamefont{Ward, Budai, Gai,
  Tischler, Yin, and Shen}}]{Ward09}
\bibinfo{author}{\bibnamefont{Ward}, \bibfnamefont{T.~Z.}},
  \bibinfo{author}{\bibnamefont{Budai}, \bibfnamefont{J.~D.}},
  \bibinfo{author}{\bibnamefont{Gai}, \bibfnamefont{Z.}},
  \bibinfo{author}{\bibnamefont{Tischler}, \bibfnamefont{J.~Z.}},
  \bibinfo{author}{\bibnamefont{Yin}, \bibfnamefont{L.}} \bibnamefont{\&}
  \bibinfo{author}{\bibnamefont{Shen}, \bibfnamefont{J.}}
  \bibinfo{title}{Elastically driven anisotropic percolation in electronic phase-separated manganites}.
  \bibinfo{journal}{\textit{Nature Phys.}} \textbf{\bibinfo{volume}{5}},
  \bibinfo{pages}{885-888} (\bibinfo{year}{2009}).

\bibitem[{\citenamefont{Tokura and Nagaosa}(2000)}]{Tokura00}
\bibinfo{author}{\bibnamefont{Tokura}, \bibfnamefont{Y.}} \bibnamefont{\&}
  \bibinfo{author}{\bibnamefont{Nagaosa}, \bibfnamefont{N.}}
  \bibinfo{title}{Orbital physics in transition-metal oxides}.
  \bibinfo{journal}{\textit{Science}} \textbf{\bibinfo{volume}{288}},
  \bibinfo{pages}{462-468} (\bibinfo{year}{2000}).

\bibitem[{\citenamefont{Schlom et~al.}(2007)\citenamefont{Schlom, Chen, Eom,
  Rabe, Streiffer, and Triscone}}]{Schlom07}
\bibinfo{author}{\bibnamefont{Schlom}, \bibfnamefont{D.~G.}},
  \bibinfo{author}{\bibnamefont{Chen}, \bibfnamefont{L.~Q.}},
  \bibinfo{author}{\bibnamefont{Eom}, \bibfnamefont{C.~B.}},
  \bibinfo{author}{\bibnamefont{Rabe}, \bibfnamefont{K.~M.}},
  \bibinfo{author}{\bibnamefont{Streiffer}, \bibfnamefont{S.~K.}}
  \bibnamefont{\&} \bibinfo{author}{\bibnamefont{Triscone}, \bibfnamefont{J.~M.}}
  \bibinfo{title}{Strain tuning of ferroelectric thin films}.
  \bibinfo{journal}{\textit{Ann. Rev. Mater. Res.}}
  \textbf{\bibinfo{volume}{37}}, \bibinfo{pages}{589-626} (\bibinfo{year}{2007}).

\bibitem[{\citenamefont{Ramesh and Spaldin}(2007)}]{Ramesh07}
\bibinfo{author}{\bibnamefont{Ramesh}, \bibfnamefont{R.}} \bibnamefont{\&}
  \bibinfo{author}{\bibnamefont{Spaldin}, \bibfnamefont{N.~A.}}
  \bibinfo{title}{Multiferroics: Progress and prospects in thin films}.
  \bibinfo{journal}{\textit{Nature Mater.}} \textbf{\bibinfo{volume}{6}},
  \bibinfo{pages}{21-29} (\bibinfo{year}{2007}).

\bibitem[{\citenamefont{He et~al.}(2004)\citenamefont{He, Wells, Ban, Alpay,
  Grenier, Shapiro, Weidong, Clark, and Xi}}]{He04}
\bibinfo{author}{\bibnamefont{He}, \bibfnamefont{F.}},
  \bibinfo{author}{\bibnamefont{Wells}, \bibfnamefont{B.~O.}},
  \bibinfo{author}{\bibnamefont{Ban}, \bibfnamefont{Z.-G.}},
  \bibinfo{author}{\bibnamefont{Alpay}, \bibfnamefont{S.-P.}},
  \bibinfo{author}{\bibnamefont{Grenier}, \bibfnamefont{S.~M.}},
  \bibinfo{author}{\bibnamefont{Shapiro}, \bibfnamefont{S.~M.}},
  \bibinfo{author}{\bibnamefont{Weidong}, \bibfnamefont{S.}},
  \bibinfo{author}{\bibnamefont{Clark}, \bibfnamefont{A.}} \bibnamefont{\&}
  \bibinfo{author}{\bibnamefont{Xi}, \bibfnamefont{X.~X.}}
  \bibinfo{title}{Structural phase transitions in epitaxial perovskite films}.
  \bibinfo{journal}{\textit{Phys.\ Rev.\ B}} \textbf{\bibinfo{volume}{70}},
  \bibinfo{pages}{235405} (\bibinfo{year}{2004}).

\bibitem[{\citenamefont{Zeches et~al.}(2009)\citenamefont{Zeches, Rossell,
  Zhang, Hatt, He, Yang, Kumar, Wang, Melville, Adamo et~al.}}]{Ramesh09}
\bibinfo{author}{\bibnamefont{Zeches}, \bibfnamefont{R.~J.}},
  \bibinfo{author}{\bibnamefont{Rossell}, \bibfnamefont{M.~D.}},
  \bibinfo{author}{\bibnamefont{Zhang}, \bibfnamefont{J.~X.}},
  \bibinfo{author}{\bibnamefont{Hatt}, \bibfnamefont{A.~J.}},
  \bibinfo{author}{\bibnamefont{He}, \bibfnamefont{Q.}},
  \bibinfo{author}{\bibnamefont{Yang}, \bibfnamefont{C.-H.}},
  \bibinfo{author}{\bibnamefont{Kumar}, \bibfnamefont{A.}},
  \bibinfo{author}{\bibnamefont{Wang}, \bibfnamefont{C.~H.}},
  \bibinfo{author}{\bibnamefont{Melville}, \bibfnamefont{A.}},
  \bibinfo{author}{\bibnamefont{Adamo}, \bibfnamefont{C.}},
  \bibnamefont{et~al.}.
  \bibinfo{title}{A strain-driven morphotropic phase boundary in BiFeO$_3$}.
  \bibinfo{journal}{\textit{Science}}
  \textbf{\bibinfo{volume}{326}}, \bibinfo{pages}{977-980} (\bibinfo{year}{2009}).

\bibitem[{\citenamefont{Torrance et~al.}(1992)\citenamefont{Torrance, Lacorre,
  Nazzal, Ansaldo, and Niedermayer}}]{Torrance92}
\bibinfo{author}{\bibnamefont{Torrance}, \bibfnamefont{J.~B.}},
  \bibinfo{author}{\bibnamefont{Lacorre}, \bibfnamefont{P.}},
  \bibinfo{author}{\bibnamefont{Nazzal}, \bibfnamefont{A.~I.}},
  \bibinfo{author}{\bibnamefont{Ansaldo}\bibfnamefont{E.~J.}}
  \bibnamefont{\&}
  \bibinfo{author}{\bibnamefont{Niedermayer}, \bibfnamefont{C.}}
  \bibinfo{title}{Systematic study of insulator-metal transitions in perovskites $R$NiO$_3$ ($R$=Pr,Nd,Sm,Eu) due to closing of charge-transfer gap}.
  \bibinfo{journal}{\textit{Phys. Rev. B}} \textbf{\bibinfo{volume}{45}},
  \bibinfo{pages}{8209-8212} (\bibinfo{year}{1992}).

\bibitem[{\citenamefont{Chmaissem et~al.}(2001)\citenamefont{Chmaissem,
  Dabrowski, Kolesnik, Mais, Brown, Kruk, Prior, Pyles, and
  Jorgensen}}]{Chmaissem01}
\bibinfo{author}{\bibnamefont{Chmaissem}, \bibfnamefont{O.}},
  \bibinfo{author}{\bibnamefont{Dabrowski}, \bibfnamefont{B.}},
  \bibinfo{author}{\bibnamefont{Kolesnik}, \bibfnamefont{S.}},
  \bibinfo{author}{\bibnamefont{Mais}, \bibfnamefont{J.}},
  \bibinfo{author}{\bibnamefont{Brown}, \bibfnamefont{D.~E.}},
  \bibinfo{author}{\bibnamefont{Kruk}, \bibfnamefont{R.}},
  \bibinfo{author}{\bibnamefont{Prior}, \bibfnamefont{P.}},
  \bibinfo{author}{\bibnamefont{Pyles}, \bibfnamefont{B.}} \bibnamefont{\&}
  \bibinfo{author}{\bibnamefont{Jorgensen}, \bibfnamefont{J.~D.}}
  \bibinfo{title}{Relationship between structural parameters and the N\'eel temperature in Sr$_{1-x}$Ca$_x$MnO$_3$ (0 $\leq x \leq$ 1) and Sr$_{1-y}$Ba$_y$MnO$_3$ ($y \leq 0.2)$}.
  \bibinfo{journal}{\textit{Phys. Rev. B}} \textbf{\bibinfo{volume}{64}},
  \bibinfo{pages}{134412} (\bibinfo{year}{2001}).

\bibitem[{\citenamefont{Mizokawa et~al.}(1999)\citenamefont{Mizokawa, Khomskii,
  and Sawatzky}}]{Mizokawa99}
\bibinfo{author}{\bibnamefont{Mizokawa}, \bibfnamefont{T.}},
  \bibinfo{author}{\bibnamefont{Khomskii}, \bibfnamefont{D.~I.}}
  \bibnamefont{\&} \bibinfo{author}{\bibnamefont{Sawatzky}, \bibfnamefont{G.~A.}}
  \bibinfo{title}{Interplay between orbital ordering and lattice distortions in LaMnO$_3$, YVO$_3$, and YTiO$_3$}.
  \bibinfo{journal}{\textit{Phys. Rev. B}}
  \textbf{\bibinfo{volume}{60}}, \bibinfo{pages}{7309-7313} (\bibinfo{year}{1999}).

\bibitem[{\citenamefont{Choi et~al.}(2004)\citenamefont{Choi, Biegalski, Li,
  Sharan, Schubert, Uecker, Reiche, Chen, Pan, Gopalan et~al.}}]{Choi04}
\bibinfo{author}{\bibnamefont{Choi}, \bibfnamefont{K.~J.}},
  \bibinfo{author}{\bibnamefont{Biegalski}, \bibfnamefont{M.}},
  \bibinfo{author}{\bibnamefont{Li}, \bibfnamefont{Y.~L.}},
  \bibinfo{author}{\bibnamefont{Sharan}, \bibfnamefont{A.}},
  \bibinfo{author}{\bibnamefont{Schubert}, \bibfnamefont{J.}},
  \bibinfo{author}{\bibnamefont{Uecker}, \bibfnamefont{R.}},
  \bibinfo{author}{\bibnamefont{Reiche}, \bibfnamefont{P.}},
  \bibinfo{author}{\bibnamefont{Chen}, \bibfnamefont{Y.~B.}},
  \bibinfo{author}{\bibnamefont{Pan}, \bibfnamefont{X.~Q.}},
  \bibinfo{author}{\bibnamefont{Gopalan}, \bibfnamefont{V.}},
  \bibinfo{author}{\bibnamefont{Chen}, \bibfnamefont{L.-Q.}},
  \bibinfo{author}{\bibnamefont{Schlom}, \bibfnamefont{D. G.}}
  \bibnamefont{\&} \bibinfo{author}{\bibnamefont{Eom}, \bibfnamefont{C. B.}}
  \bibinfo{title}{Enhancement of ferroelectricity in strained BaTiO$_3$ thin films}.
  \bibinfo{journal}{\textit{Science}}
  \textbf{\bibinfo{volume}{306}}, \bibinfo{pages}{1005-1009} (\bibinfo{year}{2004}).

\bibitem[{\citenamefont{Haeni et~al.}(2004)\citenamefont{Haeni, Irvin, Chang,
  Uecker, Reiche, Li, Choudhury, Tian, Hawley, Craigo et~al.}}]{Haeni04}
\bibinfo{author}{\bibnamefont{Haeni}, \bibfnamefont{J.~H.}},
  \bibinfo{author}{\bibnamefont{Irvin}, \bibfnamefont{P.}},
  \bibinfo{author}{\bibnamefont{Chang}, \bibfnamefont{W.}},
  \bibinfo{author}{\bibnamefont{Uecker}, \bibfnamefont{R.}},
  \bibinfo{author}{\bibnamefont{Reiche}, \bibfnamefont{P.}},
  \bibinfo{author}{\bibnamefont{Li}, \bibfnamefont{Y.~L.}},
  \bibinfo{author}{\bibnamefont{Choudhury}, \bibfnamefont{S.}},
  \bibinfo{author}{\bibnamefont{Tian}, \bibfnamefont{W.}},
  \bibinfo{author}{\bibnamefont{Hawley}, \bibfnamefont{M.~E.}},
  \bibinfo{author}{\bibnamefont{Craigo}, \bibfnamefont{B.}},
  \bibnamefont{et~al.},
  \bibinfo{title}{Room-temperature ferroelectricity in strained SrTiO$_3$}.
  \bibinfo{journal}{\textit{Nature}}
  \textbf{\bibinfo{volume}{430}}, \bibinfo{pages}{758-761} (\bibinfo{year}{2004}).

\bibitem[{\citenamefont{Zayak et~al.}(2006)\citenamefont{Zayak, Huang, Neaton,
  and Rabe}}]{Zayak06}
\bibinfo{author}{\bibnamefont{Zayak}, \bibfnamefont{A.~T.}},
  \bibinfo{author}{\bibnamefont{Huang}, \bibfnamefont{X.}},
  \bibinfo{author}{\bibnamefont{Neaton}, \bibfnamefont{J.~B.}}
  \bibnamefont{\&} \bibinfo{author}{\bibnamefont{Rabe}, \bibfnamefont{K.~M.}}
  \bibinfo{title}{Structural, electronic, and magnetic properties of SrRuO$_3$ under epitaxial strain}.
  \bibinfo{journal}{\textit{Phys. Rev. B}} \textbf{\bibinfo{volume}{74}},
  \bibinfo{eid}{094104} (\bibinfo{year}{2006}).

\bibitem[{\citenamefont{Hatt and Spaldin}(2008)}]{Hatt08}
\bibinfo{author}{\bibnamefont{Hatt}, \bibfnamefont{A.~J.}} \bibnamefont{\&}
  \bibinfo{author}{\bibnamefont{Spaldin}, \bibfnamefont{N.~A.}}
  \bibinfo{title}{Competition between polar distortions and octahedral rotations in epitaxially strained LaAlO$_3$}.
  (\bibinfo{year}{2008}), \urlprefix\url{arXiv:0808.3792 (unpublished)}.

\bibitem[{\citenamefont{Xie et~al.}(2008)\citenamefont{Xie, Budnick, Hines,
  Wells, and Woicik}}]{Xie08}
\bibinfo{author}{\bibnamefont{Xie}, \bibfnamefont{C.~K.}},
  \bibinfo{author}{\bibnamefont{Budnick}, \bibfnamefont{J.~I.}},
  \bibinfo{author}{\bibnamefont{Hines}, \bibfnamefont{W.~A.}},
  \bibinfo{author}{\bibnamefont{Wells}, \bibfnamefont{B.~O.}} \bibnamefont{\&}
  \bibinfo{author}{\bibnamefont{Woicik}, \bibfnamefont{J.~C.}},
  \bibinfo{title}{Strain-induced change in local structure and its effect on the ferromagnetic properties of La$_{0.5}$Sr$_{0.5}$CoO$_3$ thin films}.
  \bibinfo{journal}{\textit{Appl. Phys. Lett.}} \textbf{\bibinfo{volume}{93}},
  \bibinfo{pages}{182507} (\bibinfo{year}{2008}).

\bibitem[{\citenamefont{Woodward and Reaney}(2005)}]{Woodward05}
\bibinfo{author}{\bibnamefont{Woodward}, \bibfnamefont{D.~I.}} \bibnamefont{\&}
  \bibinfo{author}{\bibnamefont{Reaney}, \bibfnamefont{I.~M.}},
  \bibinfo{title}{Electron diffraction of tilted perovskites}.
  \bibinfo{journal}{\textit{Acta Cryst. B}} \textbf{\bibinfo{volume}{61}},
  \bibinfo{pages}{387-399} (\bibinfo{year}{2005}).

\bibitem[{\citenamefont{Jia et~al.}(2009)\citenamefont{Jia, Mi, Faley, Poppe,
  Schubert, and Urban}}]{Jia09}
\bibinfo{author}{\bibnamefont{Jia}, \bibfnamefont{C.~L.}},
  \bibinfo{author}{\bibnamefont{Mi}, \bibfnamefont{S.~B.}},
  \bibinfo{author}{\bibnamefont{Faley}, \bibfnamefont{M.}},
  \bibinfo{author}{\bibnamefont{Poppe}, \bibfnamefont{U.}},
  \bibinfo{author}{\bibnamefont{Schubert}, \bibfnamefont{J.}} \bibnamefont{\&}
  \bibinfo{author}{\bibnamefont{Urban}, \bibfnamefont{K.}},
  \bibinfo{title}{Oxygen octahedron reconstruction in the SrTiO$_3$/LaAlO$_3$ heterointerfaces investigated using aberration-corrected ultrahigh-resolution transmission electron microscopy}.
  \bibinfo{journal}{\textit{Phys. Rev. B}} \textbf{\bibinfo{volume}{79}},
  \bibinfo{eid}{081405(R)} (\bibinfo{year}{2009}).

\bibitem[{\citenamefont{Glazer}(1975)}]{Glazer75}
\bibinfo{author}{\bibnamefont{Glazer}, \bibfnamefont{A.~M.}}
\bibinfo{title}{Simple ways of determining perovskite structures}.
  \bibinfo{journal}{\textit{Acta Cryst. A}} \textbf{\bibinfo{volume}{31}},
  \bibinfo{pages}{756-762} (\bibinfo{year}{1975}).

\bibitem[{\citenamefont{Glazer}(1972)}]{Glazer72}
\bibinfo{author}{\bibnamefont{Glazer}, \bibfnamefont{A.~M.}}
\bibinfo{title}{The classification of tilted octahedra in perovskites}.
  \bibinfo{journal}{\textit{Acta Cryst. B}} \textbf{\bibinfo{volume}{28}},
  \bibinfo{pages}{3384-3392} (\bibinfo{year}{1972}).

\bibitem[{\citenamefont{Woodward}(1997{\natexlab{a}})}]{Woodward97a}
\bibinfo{author}{\bibnamefont{Woodward}, \bibfnamefont{P.~M.}}
\bibinfo{title}{Octahedral tilting in perovskites. I. Geometrical considerations}.
  \bibinfo{journal}{\textit{Acta Cryst. B}} \textbf{\bibinfo{volume}{53}},
  \bibinfo{pages}{32-43} (\bibinfo{year}{1997}{\natexlab{a}}).

\bibitem[{\citenamefont{Woodward}(1997{\natexlab{b}})}]{Woodward97b}
\bibinfo{author}{\bibnamefont{Woodward}, \bibfnamefont{P.~M.}}
\bibinfo{title}{Octahedral tilting in perovskites. II. Structure stabilizing forces}.
  \bibinfo{journal}{\textit{Acta Cryst. B}} \textbf{\bibinfo{volume}{53}},
  \bibinfo{pages}{44-66} (\bibinfo{year}{1997}{\natexlab{b}}).

\bibitem[{\citenamefont{He et~al.}(2005)\citenamefont{He, Wells, and
  Shapiro}}]{He05}
\bibinfo{author}{\bibnamefont{He}, \bibfnamefont{F.}},
  \bibinfo{author}{\bibnamefont{Wells}, \bibfnamefont{B.~O.}} \bibnamefont{\&}
  \bibinfo{author}{\bibnamefont{Shapiro}, \bibfnamefont{S.~M.}}
  \bibinfo{title}{Strain phase diagram and domain orientation in SrTiO$_3$ thin films}.
  \bibinfo{journal}{\textit{Phys. Rev. Lett.}} \textbf{\bibinfo{volume}{94}},
  \bibinfo{pages}{176101} (\bibinfo{year}{2005}).

\bibitem[{\citenamefont{Garc\'ia-Mu\~noz
  et~al.}(1992)\citenamefont{Garc\'ia-Mu\~noz, Rodr\'iguez-Carvajal, Lacorre,
  and Torrance}}]{Garcia92}
\bibinfo{author}{\bibnamefont{Garc\'ia-Mu\~noz}, \bibfnamefont{J.~L.}},
  \bibinfo{author}{\bibnamefont{Rodr\'iguez-Carvajal}, \bibfnamefont{J.}},
  \bibinfo{author}{\bibnamefont{Lacorre}, \bibfnamefont{P.}} \bibnamefont{\&}
  \bibinfo{author}{\bibnamefont{Torrance}, \bibfnamefont{J.~B.}}
  \bibinfo{title}{Neutron-diffraction study of $R$NiO$_3$ ($R$=La,Pr,Nd,Sm): Electronically induced structural changes across the metal-insulator transition}.
  \bibinfo{journal}{\textit{Phys. Rev. B}} \textbf{\bibinfo{volume}{46}},
  \bibinfo{pages}{4414-4425} (\bibinfo{year}{1992}).

\bibitem[{\citenamefont{Scherwitzl et~al.}(2009)\citenamefont{Scherwitzl,
  Zubko, Lichtensteiger, and Triscone}}]{Scherwitzl09}
\bibinfo{author}{\bibnamefont{Scherwitzl}, \bibfnamefont{R.}},
  \bibinfo{author}{\bibnamefont{Zubko}, \bibfnamefont{P.}},
  \bibinfo{author}{\bibnamefont{Lichtensteiger}, \bibfnamefont{C.}}
  \bibnamefont{\&} \bibinfo{author}{\bibnamefont{Triscone}, \bibfnamefont{J.-M.}}
  \bibinfo{title}{Electric-field tuning of the metal-insulator transition in ultrathin films of LaNiO$_3$}.
  \bibinfo{journal}{\textit{Appl.\ Phys.\ Lett.}}
  \textbf{\bibinfo{volume}{95}}, \bibinfo{pages}{222114}
  (\bibinfo{year}{2009}).

\bibitem[{\citenamefont{Eguchi et~al.}(2009)\citenamefont{Eguchi, Chainani,
  Taguchi, Matsunami, Ishida, Horiba, Senba, Ohashi, and Shin}}]{Eguchi09}
\bibinfo{author}{\bibnamefont{Eguchi}, \bibfnamefont{R.}},
  \bibinfo{author}{\bibnamefont{Chainani}, \bibfnamefont{A.}},
  \bibinfo{author}{\bibnamefont{Taguchi}, \bibfnamefont{M.}},
  \bibinfo{author}{\bibnamefont{Matsunami}, \bibfnamefont{M.}},
  \bibinfo{author}{\bibnamefont{Ishida}, \bibfnamefont{Y.}},
  \bibinfo{author}{\bibnamefont{Horiba}, \bibfnamefont{K.}},
  \bibinfo{author}{\bibnamefont{Senba}, \bibfnamefont{Y.}},
  \bibinfo{author}{\bibnamefont{Ohashi}, \bibfnamefont{H.}} \bibnamefont{\&}
  \bibinfo{author}{\bibnamefont{Shin}, \bibfnamefont{S.}}
  \bibinfo{title}{Fermi surfaces, electron-hole asymmetry, and correlation kink in a three-dimensional Fermi liquid LaNiO$_3$}.
  \bibinfo{journal}{\textit{Phys.\ Rev.\ B}} \textbf{\bibinfo{volume}{79}},
  \bibinfo{pages}{115122} (\bibinfo{year}{2009}).

\bibitem[{\citenamefont{Hansmann et~al.}(2009)\citenamefont{Hansmann, Yang,
  Toschi, Khaliullin, Andersen, and Held}}]{Hansmann09}
\bibinfo{author}{\bibnamefont{Hansmann}, \bibfnamefont{P.}},
  \bibinfo{author}{\bibnamefont{Yang}, \bibfnamefont{X.}},
  \bibinfo{author}{\bibnamefont{Toschi}, \bibfnamefont{A.}},
  \bibinfo{author}{\bibnamefont{Khaliullin}, \bibfnamefont{G.}},
  \bibinfo{author}{\bibnamefont{Andersen}, \bibfnamefont{O.~K.}}
  \bibnamefont{\&} \bibinfo{author}{\bibnamefont{Held}, \bibfnamefont{K.}}
  \bibinfo{title}{Turning a nickelate Fermi surface into a cupratelike one through heterostructuring}.
  \bibinfo{journal}{\textit{Phys. Rev. Lett.}} \textbf{\bibinfo{volume}{103}},
  \bibinfo{pages}{016401} (\bibinfo{year}{2009}).

\bibitem[{\citenamefont{May et~al.}(2009)\citenamefont{May, Santos, and
  Bhattacharya}}]{May09}
\bibinfo{author}{\bibnamefont{May}, \bibfnamefont{S.~J.}},
  \bibinfo{author}{\bibnamefont{Santos}, \bibfnamefont{T.~S.}}
  \bibnamefont{\&}
  \bibinfo{author}{\bibnamefont{Bhattacharya}, \bibfnamefont{A.}}
  \bibinfo{title}{Onset of metallic behavior in strained (LaNiO$_3$)$_n$/(SrMnO$_3$)$_2$ superlattices}.
  \bibinfo{journal}{\textit{Phys. Rev. B}} \textbf{\bibinfo{volume}{79}},
  \bibinfo{eid}{115127} (\bibinfo{year}{2009}).

\bibitem[{\citenamefont{Hovestreydt}(1983)}]{Hovestreydt}
\bibinfo{author}{\bibnamefont{Hovestreydt}, \bibfnamefont{E.}}
\bibinfo{title}{On the atomic scattering factor for O$^{2-}$}.
  \bibinfo{journal}{\textit{Acta Cryst. A}} \textbf{\bibinfo{volume}{39}},
  \bibinfo{pages}{268-269} (\bibinfo{year}{1983}).

\bibitem[{\citenamefont{Anisimov et~al.}(1997)\citenamefont{Anisimov,
  Aryasetiawan, and Liechtenstein}}]{Anisimov97}
\bibinfo{author}{\bibnamefont{Anisimov}, \bibfnamefont{V.~I.}},
  \bibinfo{author}{\bibnamefont{Aryasetiawan}, \bibfnamefont{F.}}
  \bibnamefont{\&} \bibinfo{author}{\bibnamefont{Liechtenstein}, \bibfnamefont{A.~I.}}
  \bibinfo{title}{First-principles calculations of the electronic structure and spectra of strongly correlated systems: the {LDA+U} method}.
  \bibinfo{journal}{\textit{J.~Phys.: Condens.
  Matter}} \textbf{\bibinfo{volume}{9}}, \bibinfo{pages}{767-808}
  (\bibinfo{year}{1997}).

\bibitem[{\citenamefont{Dudarev et~al.}(1998)\citenamefont{Dudarev, Botton,
  Savrasov, Humphreys, and Sutton}}]{Dudarev98}
\bibinfo{author}{\bibnamefont{Dudarev}, \bibfnamefont{S.~L.}},
  \bibinfo{author}{\bibnamefont{Botton}, \bibfnamefont{G.~A.}},
  \bibinfo{author}{\bibnamefont{Savrasov}, \bibfnamefont{S.~Y.}},
  \bibinfo{author}{\bibnamefont{Humphreys}, \bibfnamefont{C.~J.}}
  \bibnamefont{\&} \bibinfo{author}{\bibnamefont{Sutton}, \bibfnamefont{A.~P.}}
  \bibinfo{title}{Electron-energy-loss spectra and the structural stability of nickel oxide: An LSDA+U study}.
  \bibinfo{journal}{\textit{Phys. Rev. B}}
  \textbf{\bibinfo{volume}{57}}, \bibinfo{pages}{1505-1509} (\bibinfo{year}{1998}).

\bibitem[{\citenamefont{Kresse and Furthm{\"u}ller}(1996)}]{Kresse96}
\bibinfo{author}{\bibnamefont{Kresse}, \bibfnamefont{G.}} \bibnamefont{\&}
  \bibinfo{author}{\bibnamefont{Furthm{\"u}ller}, \bibfnamefont{J.}}
  \bibinfo{title}{Efficient iterative schemes for ab initio total-energy calculations using a plane-wave basis set}.
  \bibinfo{journal}{\textit{Phys. Rev. B}} \textbf{\bibinfo{volume}{54}},
  \bibinfo{pages}{11169-11186} (\bibinfo{year}{1996}).

\bibitem[{\citenamefont{Kresse and Joubert}(1999)}]{Kresse99}
\bibinfo{author}{\bibnamefont{Kresse}, \bibfnamefont{G.}} \bibnamefont{\&}
  \bibinfo{author}{\bibnamefont{Joubert}, \bibfnamefont{D.}}
  \bibinfo{title}{From ultrasoft pseudopotentials to the projector augmented-wave method}.
  \bibinfo{journal}{\textit{Phys. Rev. B}} \textbf{\bibinfo{volume}{59}},
  \bibinfo{pages}{1758-1775} (\bibinfo{year}{1999}).

\bibitem[{\citenamefont{Stokes and Hatch}(1988)}]{Stokes/Hatch:1988}
\bibinfo{author}{\bibnamefont{Stokes}, \bibfnamefont{H.}} \bibnamefont{\&}
  \bibinfo{author}{\bibnamefont{Hatch}, \bibfnamefont{D.}}
  \emph{\bibinfo{title}{Isotropy Subgroups of the 230 Crystallographic Space
  Groups}} (\bibinfo{publisher}{World Scientific},
  \bibinfo{address}{Singapore}, \bibinfo{year}{1988}).

\bibitem[{\citenamefont{Orobengoa et~al.}(2009)\citenamefont{Orobengoa,
  Capillas, Aroyo, and Perez-Mato}}]{Orobengoa/Perez-Mato_et_al:2009}
\bibinfo{author}{\bibnamefont{Orobengoa}, \bibfnamefont{D.}},
  \bibinfo{author}{\bibnamefont{Capillas}, \bibfnamefont{C.}},
  \bibinfo{author}{\bibnamefont{Aroyo}, \bibfnamefont{M.~I.}} \bibnamefont{\&}
  \bibinfo{author}{\bibnamefont{Perez-Mato}, \bibfnamefont{J.~M.}}
  \bibinfo{title}{AMPLIMODES: Symmetry mode analysis on the Bilbao crystallographic server}.
  \bibinfo{journal}{\textit{J. Appl. Cryst.}} \textbf{\bibinfo{volume}{42}},
  \bibinfo{pages}{820-833} (\bibinfo{year}{2009}).

\bibitem[{\citenamefont{Robinson et~al.}(1971)\citenamefont{Robinson, Gibbs,
  and Ribbe}}]{Robinson71}
\bibinfo{author}{\bibnamefont{Robinson}, \bibfnamefont{K.}},
  \bibinfo{author}{\bibnamefont{Gibbs}, \bibfnamefont{G.~V.}} \bibnamefont{\&}
  \bibinfo{author}{\bibnamefont{Ribbe}, \bibfnamefont{P.~H.}}
  \bibinfo{title}{Quadratic elongation: A quantitative measure of distortion in coordination polyhedra}.
  \bibinfo{journal}{\textit{Science}} \textbf{\bibinfo{volume}{172}},
  \bibinfo{pages}{567-570} (\bibinfo{year}{1971}).

\bibitem[{\citenamefont{Christy}(1995)}]{Christy:95}
\bibinfo{author}{\bibnamefont{Christy}, \bibfnamefont{A.}}
\bibinfo{title}{Isosymmetric structural phase transitions: Phenomenology and examples}.
  \bibinfo{journal}{\textit{Acta. Cryst. B}} \textbf{\bibinfo{volume}{51}},
  \bibinfo{pages}{753-757} (\bibinfo{year}{1995}).

  \bibitem[{\citenamefont{Bleochl}(1994)}]{Bleochl94}
\bibinfo{author}{\bibnamefont{Bleochl}, \bibfnamefont{P. E.}}
  \bibinfo{title}{Projector augmented-wave method}.
  \bibinfo{journal}{\textit{Phys. Rev. B}} \textbf{\bibinfo{volume}{50}},
  \bibinfo{pages}{17953-17979} (\bibinfo{year}{1994}).

  \bibitem[{\citenamefont{Monkhorst and Pack}(1976)}]{Monkhorst76}
\bibinfo{author}{\bibnamefont{Monkhorst}, \bibfnamefont{H. L.}} \bibnamefont{\&}
  \bibinfo{author}{\bibnamefont{Pack}, \bibfnamefont{J. D.}}
  \bibinfo{title}{Special points for Brillouin-zone integrations}.
  \bibinfo{journal}{\textit{Phys. Rev. B}} \textbf{\bibinfo{volume}{13}},
  \bibinfo{pages}{5188-5192} (\bibinfo{year}{1976}).

    \bibitem[{\citenamefont{Sreedhar}(1992)}]{Sreedhar92}
\bibinfo{author}{\bibnamefont{Sreedhar}, \bibfnamefont{K.}},
\bibinfo{author}{\bibnamefont{Honig}, \bibfnamefont{J. M.}},
\bibinfo{author}{\bibnamefont{Darwin}, \bibfnamefont{M.}},
\bibinfo{author}{\bibnamefont{McElfresh}, \bibfnamefont{M.}},
\bibinfo{author}{\bibnamefont{Shand}, \bibfnamefont{P. M.}},
\bibinfo{author}{\bibnamefont{Xu}, \bibfnamefont{J.}},
\bibinfo{author}{\bibnamefont{Crooker}, \bibfnamefont{B. C.}} \bibnamefont{\&}
  \bibinfo{author}{\bibnamefont{Spalek}, \bibfnamefont{J.}}
  \bibinfo{title}{Electronic properties of the metallic perovskite LaNiO$_3$: Correlated behavior of 3$d$ electrons}.
  \bibinfo{journal}{\textit{Phys. Rev. B}} \textbf{\bibinfo{volume}{46}},
  \bibinfo{pages}{6382-6386} (\bibinfo{year}{1992}).

\end{thebibliography}
%\bibliographystyle{nature}

\newpage
\section*{Supplemental Materials}

\begin{figure}
\includegraphics[width=5.0in]{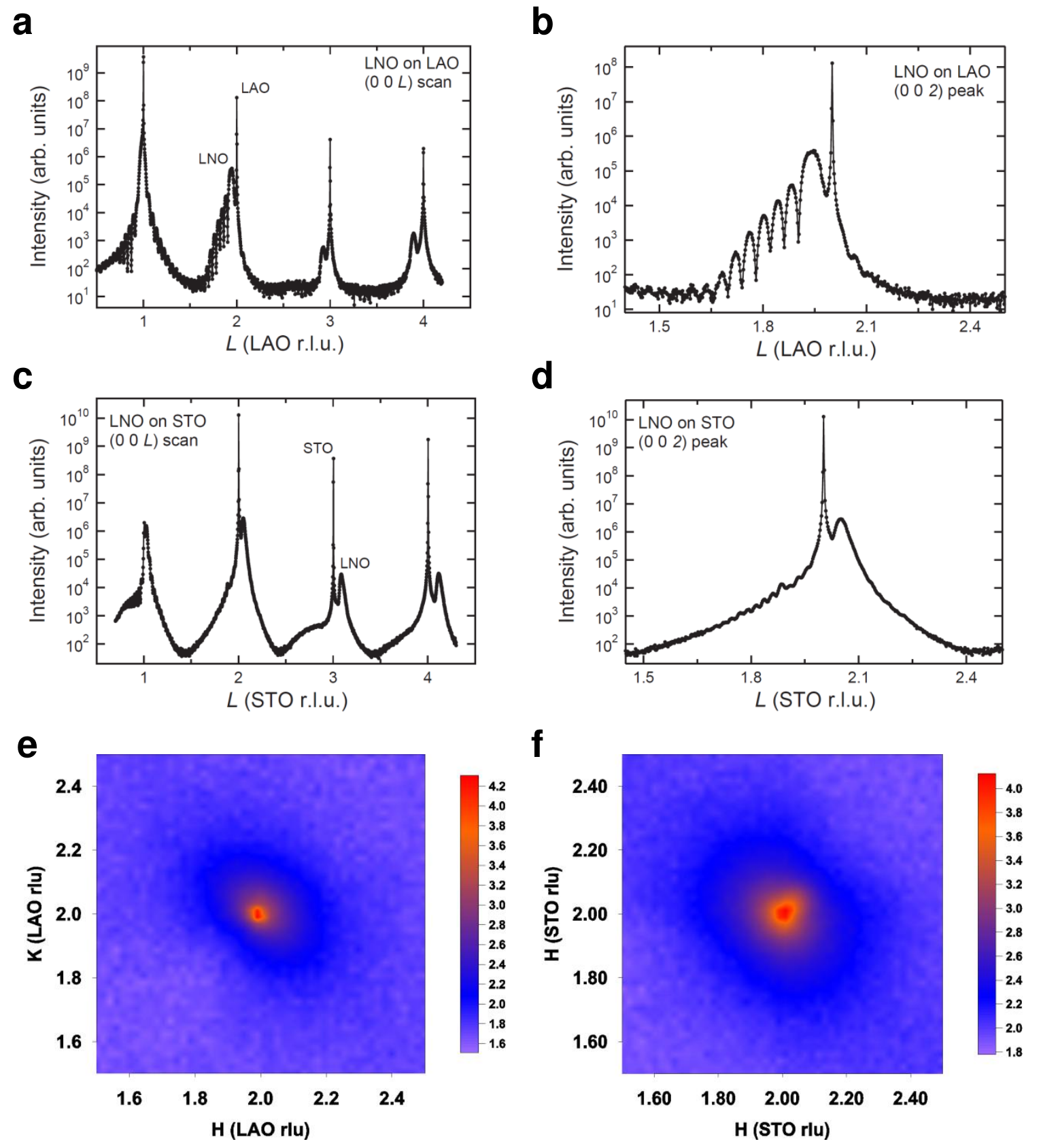}
\caption{X-ray diffraction used to determine the pseudocubic lattice parameters of the LNO films.  \textbf{a,b} Extended (0 0 $L$) scan and data near the (0 0 2) peak are used to determine the $c$-axis parameter of the LNO film grown on LAO.  \textbf{c,d} Extended (0 0 $L$) scan and data near the (0 0 2) peak are used to determine the c-axis parameter of the LNO film grown on STO.  \textbf{e,} Reciprocal space maps taken near the (2 2 1.946) condition confirm that the film grown on LAO is strained.  \textbf{f,} Reciprocal space maps taken near the (2 2 2.05) condition confirm that the film grown on STO is strained.  In all cases, the reciprocal lattice units are set to those of the substrate.}
\label{fig:S1}
\end{figure}

\textbf{Experimental results}\\
The lattice parameters of the LaNiO$_3$ (LNO) films are determined from x-ray diffraction carried out at room temperature.  Scans along the (0 0 $L$), shown in Fig.\ \ref{fig:S1}a-d, are used to determine the $c$-axis parameters of LNO on LaAlO$_3$ (LAO) and SrTiO$_3$ (STO).  The in-plane parameters are determined from $HK$ reciprocal space maps taken about the (2 2 2)film peaks (Fig.\ \ref{fig:S1}e,f).  In both LNO on STO and LAO, the films are found to be coherently strained to the substrates, exhibiting the same $a$,$b$ parameters as the substrates.

\begin{figure}
\includegraphics[width=5.5in]{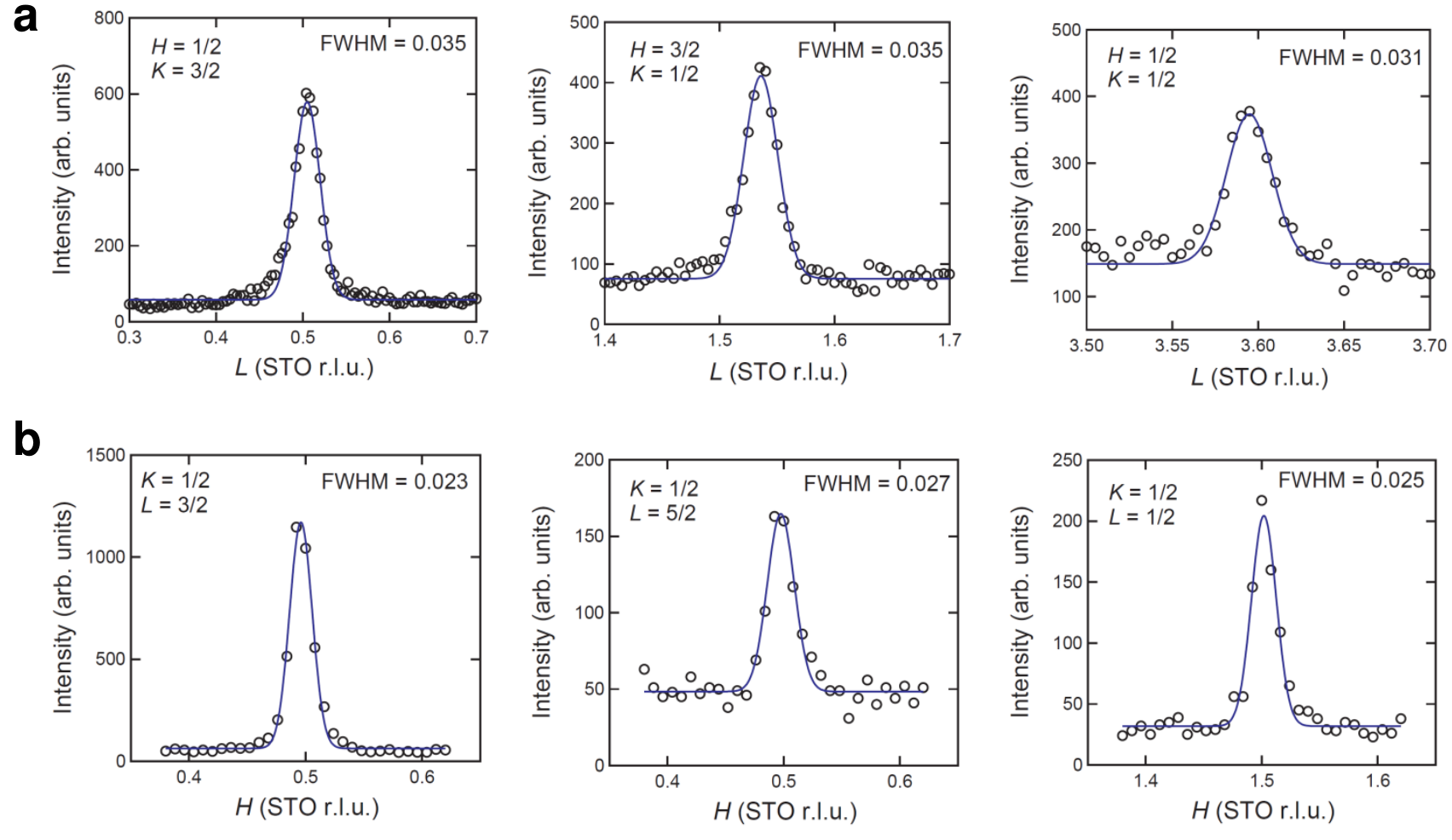}
\caption{Scans of various half-order peaks through $L$ and $H$.  The peak widths of all the $L$ scans (\textbf{a}) are roughly equal, while the widths of all the $H$ scans (\textbf{b}) are roughly equal.    From this we conclude that the half-order reciprocal lattice points all have approximately the same width along $H$, $K$, and $L$.}
\label{fig:S2}
\end{figure}

Half-order peaks, arising from the octahedral rotations, are measured to determine the oxygen positions.  Figure \ref{fig:S2} shows scans through various half-order peaks along the $H$ and $L$ directions, confirming that the peak widths are independent on their location on the reciprocal lattices.

\begin{figure}
\includegraphics[width=5.0in]{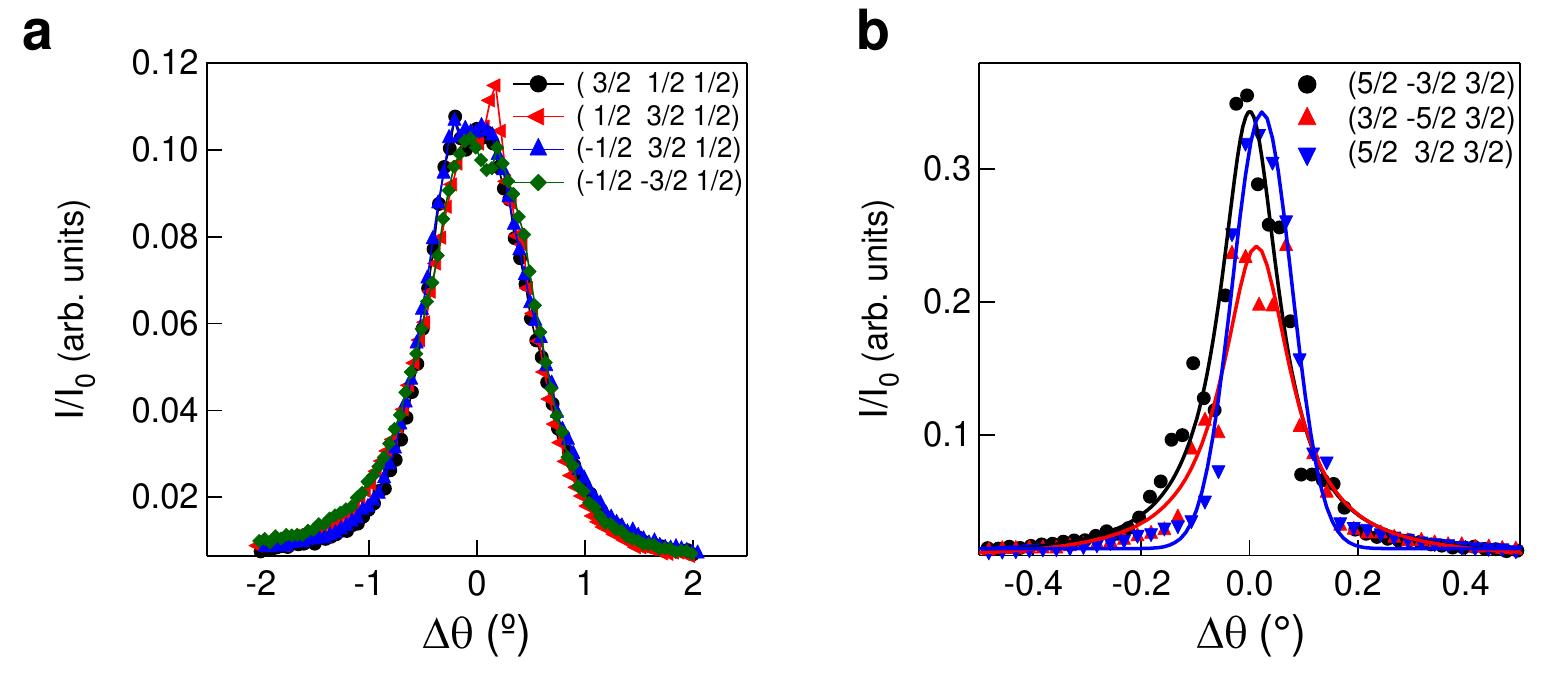}
\caption{Evidence for domains in LNO films.  Half-order diffraction peaks from the LNO grown on STO (\textbf{a}) and LAO (\textbf{b}).  The equal intensity of the (3/2 1/2 1/2)-family of peaks in LNO/STO indicate an equal population of tilt domains.  The unequal intensity of the (5/2 3/2 3/2)-family of peaks in LNO/LAO indicate that some domains are preferred over others.  The (1/2 1/2 5/2)-family of peaks also exhibits unequal intensities.}
\label{fig:S3}
\end{figure}

\begin{figure}
\includegraphics[width=5.0in]{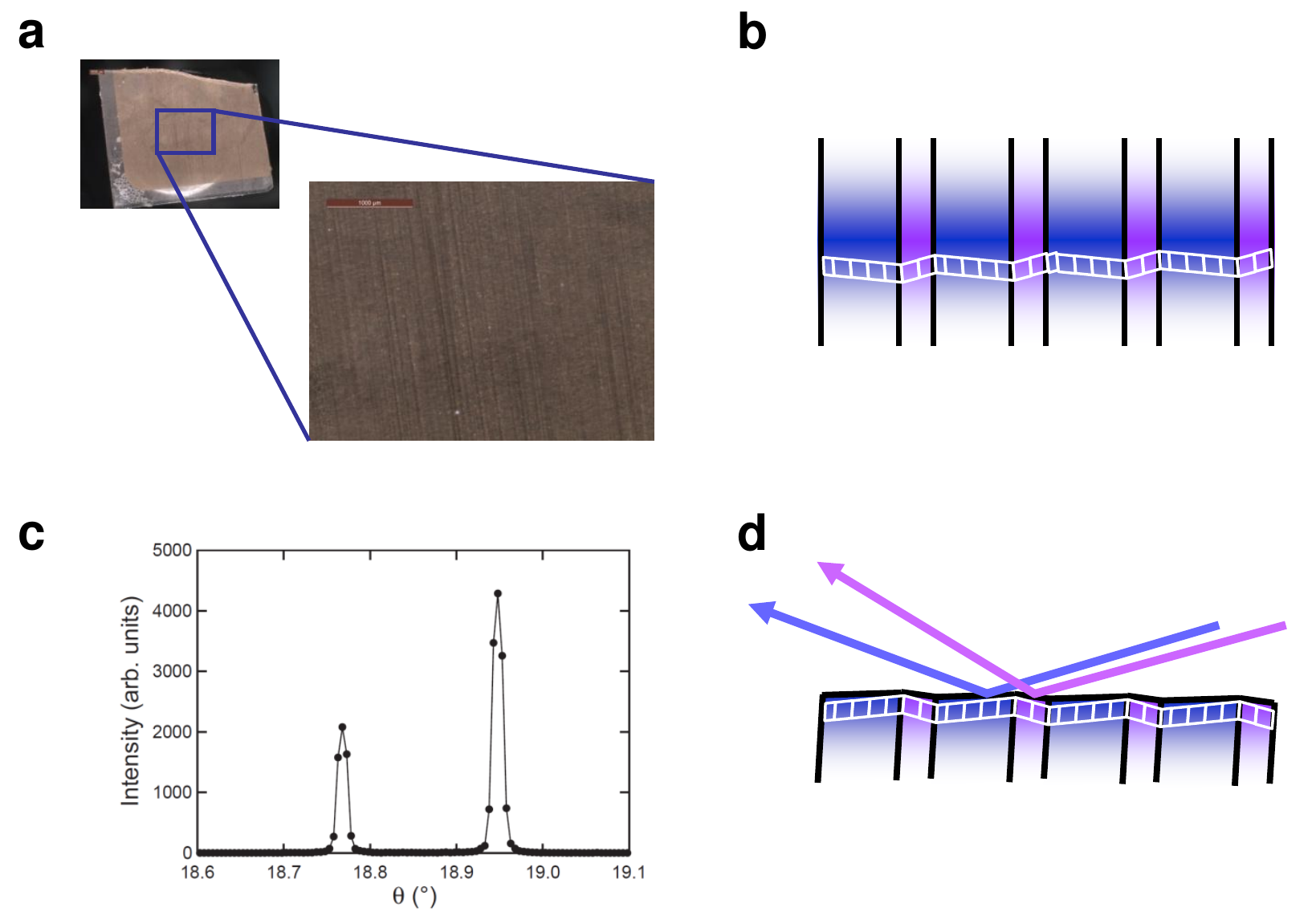}
\caption{Twinning in LAO substates.  \textbf{a,} An optical micrograph of the LNO/LAO sample.  Twin boundaries run parallel along one of the $a$-axis of the LAO substrate.  \textbf{b,} Schematic view of the twin structure of the LAO substrate and the perovskite unit cells. \textbf{c,} The (0 0 2)$_{perovskite}$ rocking curve of LAO substrate reveals the presence of two dominant orientations . The intensity ratio of the two peaks indicates the ratio of the two twin domains is approximately 1:2.  \textbf{d,} The reflection conditions of the twin domains.  The rhombohedral tilting angle is exaggerated for clarity.}
\label{fig:S4}
\end{figure}

There are four possible domains that give rise to different structure factors for each peak.  The domain volume fractions are obtained by comparing the intensities of a set of symmetrically equivalent peaks with fixed $L$.  For LNO on STO, the equal intensities of the (1/2 3/2 1/2) family of peaks, shown in Fig.\ \ref{fig:S3}a, indicate the four domains have equal volume fractions.  Similar scans for the LNO/LAO film (Fig.\ \ref{fig:S3}b) indicate unequal domain populations, with only two domains present in a ratio of 2:1.  The ratio of twin domains in the LAO substrate is also 2:1 (Fig.\ \ref{fig:S4}) suggesting that the rotational domains present in the LNO film are imprinted from the LAO substrate.

With the domain volume fractions determined, scans were carried out through multiple half-order Bragg peaks to determine the oxygen positions.  The bond angles and lengths are obtained and are summarized in Table S1.

\begin{table}
\begin{ruledtabular}
\begin{tabular}{cccccccc}
 & LNO/STO (XRD) & LNO/LAO (XRD) & LNO/STO (DFT) & LNO/LAO (DFT) \\
\hline
$d_{in-plane}$ (\AA) & 1.968 $\pm$ 0.002 & 1.916 $\pm$ 0.005 & 1.970 & 1.921\\
$d_{out-of-plane}$ (\AA) & 1.933 $\pm$ 0.002 & 1.949 $\pm$ 0.002 & 1.935 & 1.950\\
$\theta_{in-plane} (^{\circ})$ & 165.8 $\pm$ 0.5 & 164.0 $\pm$ 2.0 & 164.7 & 161.5\\
$\theta_{out-of-plane} (^{\circ})$ & 159.9 $\pm$ 0.6 & 175.2 $\pm$ 0.6 & 159.3 & 173.7\\
\end{tabular}
\end{ruledtabular}
\caption{\label{tab:table1}The bond angles and lengths as determined by x-ray diffraction (XRD) and density functional theory (DFT).}
\end{table}

\clearpage
\textbf{Geometric constraints on the octahedral rotation system}\\

\begin{figure}
\includegraphics[width=4.0in]{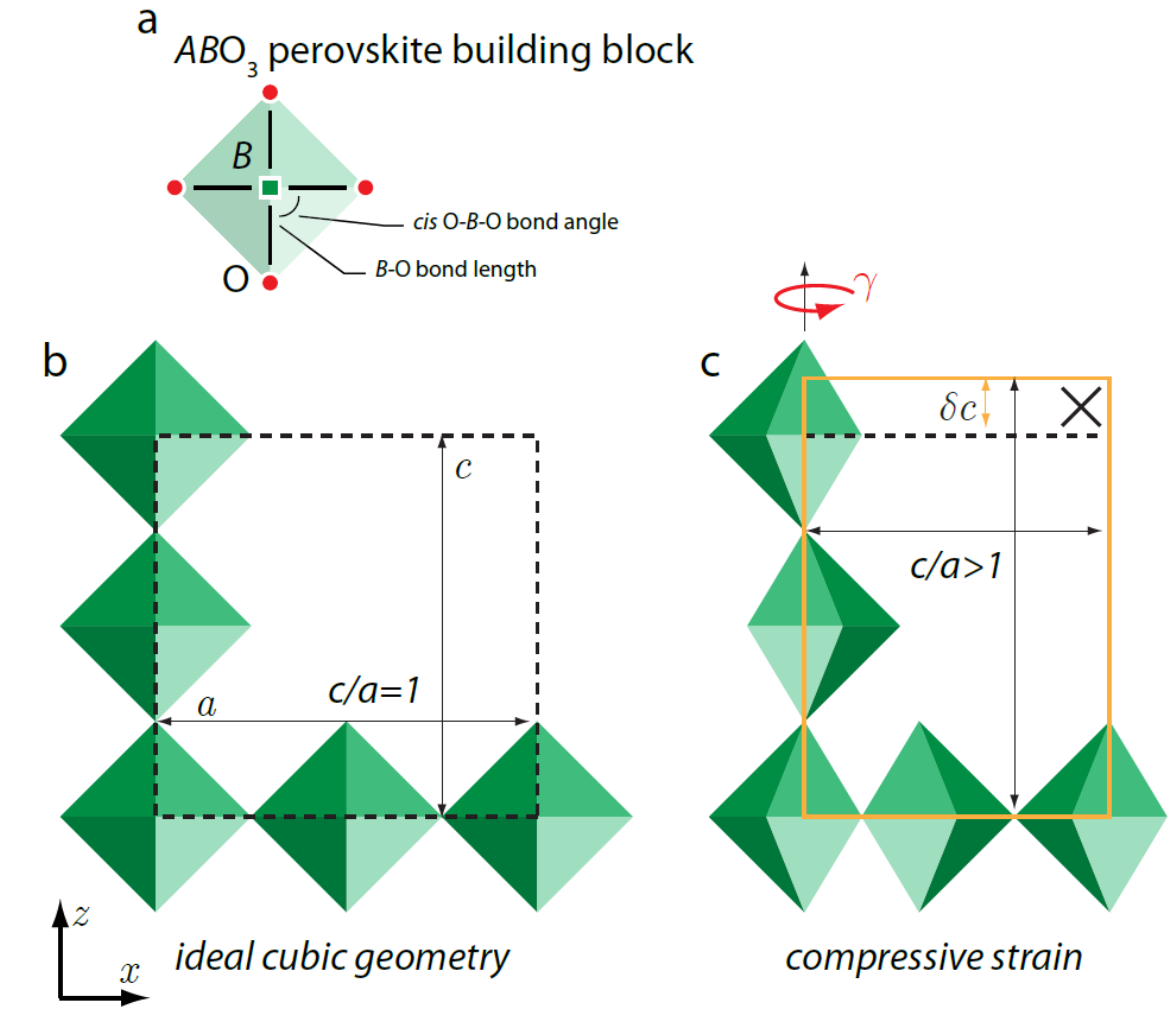}
\caption{Geometric illustration for the origin of the isosymmetric phase transition at the axial ratio of $c$/$a$=1.  \textbf{a}, The BO$_6$ octahedral building block in perovskite oxides. \textbf{b,} Ideal reference structure for a perovskite with cubic symmetry. The ideal, cornershared octahedral units, are shown to readily tile the unit cell in the $x$ and $z$-directions. \textbf{c,} The effect of bi-axial compressive strain causes the $a$-axis parameter to shrink, and therefore increases the $c$ parameter due to elasticity. If we fix the B-O bond lengths about their ideal length as in (\textbf{b}), then the octahedral must rotate by an amount $\gamma$ along the $z$-direction. The geometric mismatch $\delta c$ between the ideal and elongated unit cells then drives the octahedral to rotate about an in-plane axis, in order to preserve the original B-O bond lengths and O-B-O internal angles.}
\label{fig:S5}
\end{figure}

In Fig. \ref{fig:S5}a we show the BO$_6$ octahedral unit as the building block for the perovskite structure. We consider in Fig. \ref{fig:S5}b, an ideal cubic structure with an axial ratio $c$/$a$ defined to be unity. In this case all of the B-O bonds and O-B-O bonds are equal yielding the ideal six-coordinate octahedron.

In Fig. \ref{fig:S5}c we schematically show the effect of applying bi-axial compressive strain to a 3x3x3 supercell of the primitive (and cubic) 5-atom perovskite cell. We enforce that the B-O bond lengths and O-B-O angles in each individual octahedron remain identical. [This assumption is good for small strains, and agrees with Pauling's rule on parsimony.] In order to accommodate the change in the in-plane $a$ lattice constant, the octahedra rotate about the $c$-axis by an amount $\gamma$. In the out-of-plane direction, if the octahedra \textit{do not change} their rotational pattern, then by the requirement that the out-of-plane lattice constant $c$ increases (from elastic theory), there results a mismatch $\delta c$ between the repeat unit cell corresponding to $c$/$a$=1 (dashed line), and that with $c$/$a>1$ (yellow line). This mismatch that occurs when the octahedra along c maintain their original rotation (in this case zero) and bond lengths, requires that the magnitude of the octahedral rotation pattern about an axis parallel to the plane of the strain changes to make the repeat unit cells coincide with one another. The need to change the octahedral bond angles, then leads to the abrupt, first-order phase transition. Interestingly the origin for this isosymmetric constraint can be completely attributed to the corner connectivity of the octahedral units and their antiferrodistortive nature that is compatible with a three-fold axis but incompatible with the bi-axial constraint imposed on the film by the substrate.

\clearpage
\textbf{Electronic band structure calculations}\\

\begin{figure}
\includegraphics[width=6.0in]{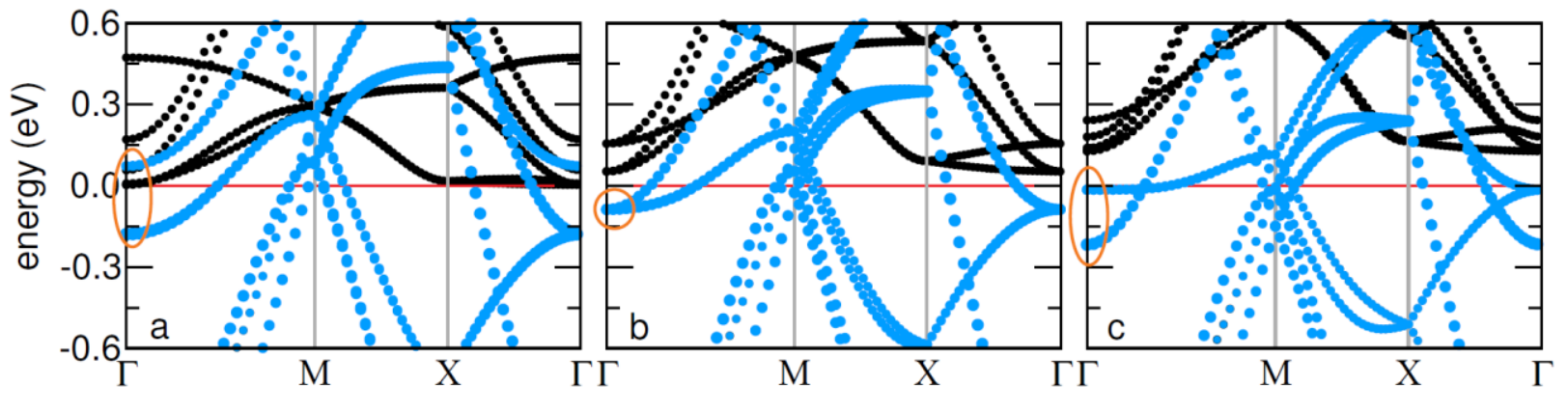}
\caption{Calculated electronic structure.  The calculated band structure of LaNiO$_3$ on LaAlO$_3$ (\textbf{a}), bulk LaNiO$_3$ (\textbf{b}), and LaNiO$_3$ on SrTiO$_3$ (\textbf{c}) along high symmetry lines in the Brillouin given as $\Gamma$(0,0,0)$\to$M(1/2,1/2,0)$\to$X(1/2,0,0)$\to$$\Gamma$(0,0,0). For each strain state, LaNiO$_3$ remains metallic. The role of strain on the electronic structure is to partially split the orbital degeneracy between the majority spin $d_{x^2}-_{y^2}$ and $d_{{3z}^2}-_{r^2}$ orbitals (blue points) as indicated at the $\Gamma$-point. The Fermi level is denoted by the solid red line at 0 eV.}
\label{fig:S6}
\end{figure}

\begin{figure}
\includegraphics[width=5.0in]{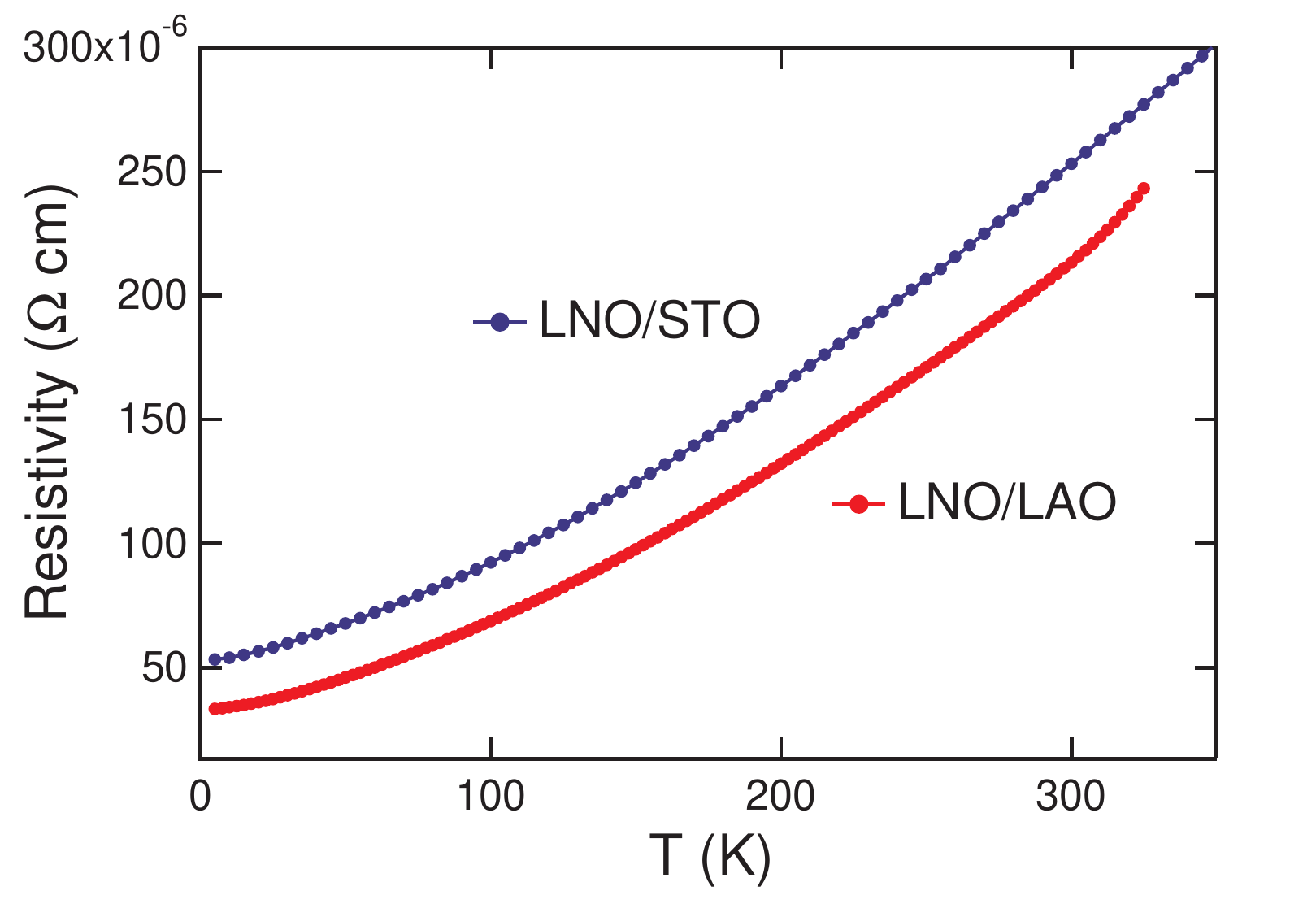}
\caption{Resistivity of the LNO films on LAO and STO.  Both films show metallic behavior with a similar temperature dependence.}
\label{fig:S7}
\end{figure}

We next plot the electronic band structures for the two strained films on STO and LAO using the experimental nickelate structures along the high symmetry lines in a non-primitive (2x2x1) unit cell of the monoclinic structure. This supercell doubles the number of bands, however, it makes the identification of the orbital contributions to the band structure more transparent. In Fig. \ref{fig:S6}a we show the calculated band structure for the experimentally determined nickelate film on the LAO substrate and compare it to the bulk experimental structure in Fig. \ref{fig:S6}b and to that on STO substrate in Fig. \ref{fig:S6}c. For each strain state the band structure is found to be fully metallic, with strain partially lifting the majority spin $e_g$ orbital degeneracy at the $\Gamma$-point around the Fermi level (set to zero here).  The metallic nature of the films is evident in the resistivity behavior, shown in Fig. \ref{fig:S7}.

\end{document}